\documentclass[a4,journal]{IEEEtran}

\ifCLASSINFOpdf
\else
\fi
\usepackage{comment}
\usepackage{multirow}
\usepackage{subcaption}
\usepackage{tabularx,booktabs}
\usepackage{graphicx}
\usepackage{color}
\usepackage{epstopdf}
\usepackage{setspace}
\usepackage{soul}
\usepackage{lettrine}
\usepackage{hyperref}
\usepackage[square,comma,sort&compress,numbers]{natbib}
\usepackage{amsmath}
\usepackage{amssymb}
\usepackage{algpseudocode,algorithm}
\graphicspath{{Figures/}}
\usepackage{multicol, blindtext}
\newtheorem{lemma}{Lemma}
\newtheorem{remark}{Remark}
\DeclareMathSizes{10}{9}{7}{6}
\IEEEaftertitletext{\vspace{-2\baselineskip}}

\begin{document}
\title {Non-orthogonal HARQ for URLLC:\\ Design and Analysis}  
\author{Faisal~Nadeem,~\IEEEmembership{Student Member,~IEEE}, Mahyar~Shirvanimoghaddam,~\IEEEmembership{Senior Member,~IEEE}, Yonghui~Li,~\IEEEmembership{Fellow,~IEEE}, Branka~Vucetic,~\IEEEmembership{Fellow,~IEEE}
\thanks{Authors are with the Centre for IoT and Telecommunications, School of Electrical and Information Engineering, The University of
Sydney, NSW 2006, Australia. (email: \{faisal.nadeem, mahyar.shm, yonghui.li, branka.vucetic\}@sydney.edu.au.)}

\thanks{This paper was presented in part at the 2019 IEEE International Communications Conference (ICC), Dublin, Ireland. This work was supported by the Australian Research Council through the Discovery Projects under Grants DP180100606 and DP190101988.}
\thanks{
This paper has been accepted for publication by IEEE Internet of Things Journal. Copyright (c) 2021 IEEE. Personal use of this material is permitted. However, permission to use this material for any other purposes must be obtained from the IEEE by sending a request to \url{pubs-permissions@ieee.org}.} }

\maketitle
\begin{abstract}
The fifth-generation (5G) of mobile standards is expected to provide ultra-reliability and low-latency communications (URLLC) for various applications and services, such as online gaming, wireless industrial control, augmented reality, and self driving cars. Meeting the contradictory requirements of URLLC, i.e., ultra-reliability and low-latency, is considered to be very challenging, especially in bandwidth-limited scenarios. Most communication strategies rely on hybrid automatic repeat request (HARQ) to improve reliability at the expense of increased packet latency due to the retransmission of failing packets. To guarantee high-reliability and very low latency simultaneously, we enhance HARQ retransmission mechanism to achieve reliability with guaranteed packet level latency and in-time delivery. The proposed non-orthogonal HARQ (N-HARQ) utilizes non-orthogonal sharing of time slots for conducting retransmission. The reliability and delay analysis of the proposed N-HARQ in the finite block length (FBL) regime shows very high performance gain in packet delivery delay over conventional HARQ in both additive white Gaussian noise (AWGN) and Rayleigh fading channels. We also propose an optimization framework to further enhance the performance of N-HARQ for single and multiple retransmission cases.
\end{abstract}
\section{Introduction}
\IEEEPARstart{T}{he} fifth-generation (5G) of mobile standards  envisions to  support not only enhanced mobile broadband (eMBB), which was conventionally the only usage scenario of mobile standards, but also  massive machine-type communications (mMTC) and  ultra-reliable low-latency communications (URLLC) \cite{holma20205g}. URLLC aims to  supports various mission critical applications, such as telesurgery, tactile Internet, factory automation (Industry 4.0), and smart grids \cite{schulz2017latency,chen2018ultra}, which require user plane latency below  1ms and reliability above 99.9999\%  for a packet of size 32 bytes with or without retransmission \cite{shirvanimoghaddam2019short}. On the other hand, mMTC mainly focuses on applications like smart metering, smart agriculture, etc.,  which involve a large number of devices mainly communicating short messages with usually relaxed delay constraints but strict energy efficiency requirements.  .

Conventional communication systems are designed to achieve high data rates, but have less requirements on latency and reliability. Usually, the packet size is long and  packet arrival deadlines are not stringent \cite{zia2014bandwidth}. This flexibility allows to repeat the failing packets for time-diversity to combat the severe channel impairments, but  it increases the latency. Achieving low latency  mandates a short packet length, but reducing the packet length will degrade the error performance \cite{shirvanimoghaddam2019short,polyanskiy2010channel}. In order to achieve ultra-reliable transmission with short packet communication, joint design of error correcting codes and retransmissions such as Hybrid automatic repeat request (HARQ)  are essential, which eventually increase latency \cite{bennis2018ultrareliable}.  Furthermore, under strict latency constraints, the availability of instantaneous channel state information (CSI) is not viable, especially at the transmitter. Therefore, low latency communication requires strategies that do not depend on instantaneous CSI availability \cite{lopez2020ultra}. In conventional adaptive modulation and coding strategies, the transmitter can adopt its modulation and channel coding scheme to maximize spectral efficiency or minimize the block error rate based on the channel knowledge. However, in delay-sensitive communication, the retransmission  mechanism such as HARQ becomes more critical due to the absence of CSI at the transmitter \cite{trillingsgaard2017generalized}.

In current cellular systems, HARQ techniques  have been widely used to achieve reliable transmission. With HARQ,
when the receiver recovers packets successfully, it feed backs the acknowledgment (ACK) signal; otherwise, it sends a negative ACK (NACK) to request a retransmission. Upon receiving a NACK, the transmitter retransmits the packet, in case of Chase combining HARQ (CC-HARQ), or send more redundancy in case of incremental redundancy HARQ (IR-HARQ). IR-HARQ performs better but requires more signaling overhead, whereas CC-HARQ has a lower complexity with slight performance degradation. The receiver combines the failing packet with its retransmissions using maximum ratio combining (MRC) or code combining for CC or IR HARQ, respectively. Performance of both IR-HARQ and CC-HARQ have been analyzed in the finite block length (FBL) regime in the additive white Gaussian noise (AWGN) channel \cite{kim2013performance}.  Delay sensitivity and reliability of HARQ in the FBL regime were also investigated for Rayleigh block fading and other communication scenarios  \cite{makki2018fast,sahin2019delay,ostman2018low}. In standard HARQ, retransmissions cause latency and queuing delay,  because waiting for the receiver's feedback and retransmission scheduling consumes extra time slots \cite{malak2018arq}.  The proactive packet dropping mechanism has been investigated to limit the queuing delay  by sacrificing the packet error rate (PER) performance  \cite{sassioui2016harq,najm2017status,sun2018optimizing}. In \cite{jabi2015multipacket}, the authors proposed a multi-packet HARQ strategy to schedule two packets simultaneously upon receiving the retransmission request. However, in \cite{sassioui2016harq,jabi2015multipacket,szczecinski2013rate}, authors considered only asymptotic block length regime and focused on improving the throughput performance  with  very little focus on latency performance. More recently, there has been some efforts in making retransmission schemes more robust for delay-sensitive communication by reducing feedback delays and better resource utilization \cite{nadas2019performance,chung2010performance,anand2018resource,dosti2017ultra}.
In the FBL regime and in particular, for URLLC, the performance of the communication system should be  analyzed at the packet level \cite{parag2012code}.  This motivates developing new strategies to achieve the target reliability under strict per-packet latency constraints \cite{bennis2018ultrareliable,shirvanimoghaddam2020dynamic,nadeem2020non}.

Recently, with the advancements of multi-user detection techniques and non-orthogonal multiple access schemes (NOMA), new directions have been considered to further enhance the performance of communication systems. These can improve radio resource utilization and have been considered key technologies for 5G \cite{liu2017non,shahab2020grant}. 
For example, in  power domain NOMA,  multiple signals are superimposed  at the transmitter, and successive interference cancellation is used to separate signals at the receiver. NOMA has been analyzed with short packet communication to meet the delay requirements of various users by efficient resource utilization \cite{xu2020latency,ghanami2020performance}.
NOMA with HARQ is investigated for its ability to enhance spectral and energy efficiency by exploiting multi-user cooperation \cite{makki2020error}.  Recently, researchers showed that NOMA effectively provides low latency and packet level delay guarantees,  especially when retransmissions are used to increase communication reliability  \cite{nadeem2020non,kotaba2020urllc}. Also in \cite{makki2020hybrid}, the authors utilized the NOMA-based retransmission, where a user is allowed to share its channel to allow retransmission packets of the other user. In this way the frequency diversity is utilized to improve retransmission reliability and delay penalty due to retransmission is shared. 
 In the aforementioned literature, mostly NOMA's benefit is highlighted for effective use of resources for  multi-user packets scheduling.  In this work, we adopt the non-orthogonal signaling between transmission of new packets and retransmissions over a single stream of packets to avoid queuing. However, the implications of this work is far reaching to both uplink and downlink and for any scenario where packet queuing is expected due to retransmissions. 

We propose a HARQ scheme for delay-sensitive communication in which retransmissions are enabled via non-orthogonal signaling to reduce  the  queuing delay. The proposed scheme, referred to as non-orthogonal HARQ (N-HARQ),  is more resilient to delay violations and simultaneously offers high reliability. More specifically, we allow  retransmissions of failing packets to share the  time slots with the new arriving  packets in a non-orthogonal fashion. For CC-HARQ the retransmissions are shared with new arriving packets in the power domain in the whole time slot. In IR-HARQ partial time slot sharing is allowed to facilitate the  variable length of redundancy. This achieves higher reliability but has higher complexity as well due to more control signaling. Major contributions of the paper are as follows:
\begin{itemize}
\item We propose a novel retransmission technique using  non-orthogonal signaling, i.e the retransmission and new packets share the same time slot, to enable HARQ retransmissions without causing queuing and excess delay. In this way, reliability is improved with a guaranteed packet arrival deadline. 
\item We develop an analytical framework to characterize the reliability and delay performance of the proposed scheme in the FBL regime in AWGN and Rayleigh fading channels. In particular, we propose a Markov model to characterize the dynamic of N-HARQ. We also analyse its performance in the fading channel by utilizing the partitioning based correlative fading channel with different states, where state transition probabilities are calculated under the FBL assumption. 
\item We propose an optimization framework to maximize throughput under the target latency constraint. The systems parameters, including the retransmission length and power-splitting ratios, are optimized to maximize the throughput at different signal-to-noise ratios (SNRs) and Doppler frequencies.
\end{itemize}
We provide extensive simulation results to investigate the performance of N-HARQ for different numbers of retransmissions, power-splitting and time-sharing parameters, packet size, and type of HARQ (IR and CC) in various channel conditions, SNRs, and Doppler frequencies. We also compare the reliability and latency performance of the proposed N-HARQ with the standard HARQ technique in both AWGN and Rayleigh fading channels to highlight its effectiveness. Results show that N-HARQ achieves the target reliability and throughput with much lower latency than standard HARQ and provide a packet-level delay guarantee. Whereas in standard HARQ, reliability is achieved at the cost of increased latency and compromised packet-level delay guarantee.

The rest of the paper is organized as follows. The system model and preliminaries on HARQ in the FBL regime are presented in Section \ref{Sec:System_Model}. In Section \ref{sec:NOMA_MODEL}, we explain the proposed N-HARQ for general retransmission. The reliability and delay performance of the proposed N-HARQ is analysed in Section \ref{sec:reliability_delay_analysis}. We provide numerical results and discussions in Section \ref{Sec:Numarical_resutls}. Finally, Section \ref{sec:conclusion} concludes the paper.

\section{System Model and Preliminaries}
\label{Sec:System_Model}
We consider a delay-sensitive application, in which a stream of $N$ packets needs to be delivered at the destination, where each packet has its delivery deadline. That is, packet $\ell$ should be delivered on or before time $T_{\ell}$. We further assume a constant packet arrival at the transmitter. We employ HARQ so that the packets that were not decoded in the first transmission round are provided another chance using retransmissions. We allow maximum $m-1$ retransmissions to limit the packet level latency ($m\ge1$). 

The channel between the transmitter and receiver, denoted by $h(t)$, is modeled by Rayleigh fading. Let $u(t)$ and $y(t)$ denote the transmitted signal and the received signal at the receiver, respectively. Then, $y(t)$ is given by:
\begin{align}
y(t)=h(t)u(t)+w(t),
\end{align}
where $w(t)\sim\mathcal{CN}(0,N_0)$ is the circularly symmetric zero-mean complex additive white Gaussian noise (AWGN) with independent real and imaginary parts, each with power spectral density $N_0/2$. The transmit signal $u(t)$ is linearly modulated and transmitted with bandwidth $B$ Hz with normalized symbol rate 1 symbol/s/Hz. We also assume that the total transmit power is $\mathbb{E}[|u(t)|^2]=P$. Following the finite-state Markov channel (FSMC) \cite{zhang1999finite}, the fading envelope process $|h(t)|\geq 0$ follows a Rayleigh distribution and the average power gain is normalized to unity, i.e. $\mathbb{E}[|h(t)|^2]=1$. We assume that the mobile terminal experiences Doppler fading due to constant relative motion. The auto-correlation function of $h(t)$ is modeled by a zeroth-order Bessel function of the first kind as $J_0(2\pi f_{\mathrm{D}}t)$, where $f_\mathrm{D}$ the Doppler frequency.
\vspace{-1.5ex}
\subsection{Time block-based FSMC}
Due to the time-varying nature of the wireless media, one of the critical challenges is the performance characterization of a delay-sensitive communication system  under correlative fading. Due to finite block lengths for a delay-sensitive communication, the overly optimistic assumption of independent and identically distributed (i.i.d.) or the same channel across packets to simplify analysis may not be practical. Due to the unavailability of CSI, the transmitter fixes the transport block (TB) size and each TB goes to a different channel state. Therefore, the finite state channel model becomes a natural choice to analyze the system with fixed TB sizes.

Similar to \cite{sahin2019delay,zhang1999finite}, we partition the fading envelop into $L$ fading states using thresholds $\boldsymbol\eta=[\eta_0,\eta_2,\ldots,\eta_{L}]$, where $\eta_1=0,\eta_{L+1}=\infty$. Let $\mathcal{L}=[1, 2, \cdots, L]$ denote the index set for $L$ fading states, where $S_{\ell}$, for $\ell\in\mathcal{L}$, denotes the $\ell$-th fading state.  At any given time, if the channel gain $|h(t)|$ lies in $[\eta_{\ell},\eta_{\ell+1}]$, the channel is said to be in state $S_{\ell}$.
We consider a block fading channel for which the channel remain constant within each codeword of $n$ symbols but varies from one block to another one \cite{zhang1999finite}. This forms a Markov chain, that is sampled at each TB.
\begin{figure}[t]
\centering
\includegraphics[width=0.65\columnwidth]{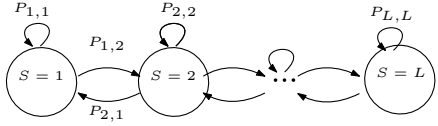}
\caption{The transport block-based finite-state Markov channel (FSMC) model of the Rayleigh fading channel.}
\label{fig:FSMM_Channel}
\end{figure}

The fading envelope partition $\eta_\ell$, the TB duration $t_{\text{TB}}$, and the Doppler frequency $f_\mathrm{D}$ are sufficient to fully characterize the underlying Markov chain. Furthermore, the complexity of the FSMC is further reduced under the assumption that the state of the FSMC only transits to adjacent states, i.e., $P_{\ell,k}=0$ whenever $|\ell-k|>1$, where $P_{\ell,k}$ is the transition probability from channel fading state $S_{\ell}$ to state $S_{k}$, where $\ell,k\in \mathcal{L}$. This assumption is valid for the finite block length transmission where the channel does not change significantly across consecutive packets. Fig. \ref{fig:FSMM_Channel} shows the state transition diagram of the finite-state MM for fading channel.  The state transition probabilities associated with the FSMC model can be calculated as follows \cite{wang1995finite}:
\begin{subequations}
\begin{align} 
P_{\ell,{\ell}+1}\left ({\boldsymbol\eta, f_{\mathrm{D}}, t_{\mathrm{TB}}}\right)&\approx\frac {N(\eta _{\ell+1}, f_{\mathrm{D}})t_{\mathrm{TB}}}{q_{\ell}(\eta_{\ell},\eta_{\ell+1})},\quad 1 \le\ell\le L-1, \label{eq:1}\\ 
P_{\ell,{\ell}-1}\left ({\boldsymbol\eta, f_{\mathrm{D}}, t_{\mathrm{TB}}}\right)&\approx\frac {N(\eta_{\ell}, f_{\mathrm{D}})t_{\mathrm{TB}}}{q_{\ell}(\eta_{\ell},\eta_{\ell+1})},\quad ~~~2\le\ell\leq L\label{eq:2}
\end{align}
\label{eq:fadingtransitions}
\end{subequations}
where $q_\ell$ is the marginal probability of channel being in state $S_{\ell}$, which is given by:
\begin{align} 
q_{\ell}= \int_{\eta_{\ell}}^{\eta_{\ell+1}} {2 x e^{-x ^{2}} dx} ,
\label{eq:marg_prob_states}
\end{align}
where $f(x)=2 x e^{-x^2}$ is the marginal probability distribution of $|h(t)|$ and $N(\eta_{\ell},f_\mathrm{D})$ is the average number of times per second that the signal envelope $|h(t)|$ crosses level $\eta_{\ell}$ under the Bessel auto-correlation function model given by \cite{wang1995finite,sadeghi2008finite}:
\begin{align} 
N(\eta_{\ell},f_{\mathrm{D}})=\sqrt{2\pi}\eta_{\ell} f_{\mathrm{D}} e^{-\eta_{\ell}^{2}}.
\label{eq:channelcrossing}
\end{align}

It can be observed from \eqref{eq:fadingtransitions} and \eqref{eq:channelcrossing} that state transition probabilities linearly increase with time block $t_\mathrm{TB}$ and $f_\mathrm{D}$. As the total outgoing probabilities of each state must be 1, $t_\mathrm{TB}$ is upper bounded by:
\begin{align}
  t_\mathrm{TB}\leq\frac{p_\ell(\eta_\ell, \eta_{\ell+1})}{N(\eta_{\ell}, f_{\mathrm{D}})+N(\eta_{\ell+1}, f_{\mathrm{D}})},\quad \forall \ell\in \mathcal{L},
  \label{eq:state_duraton}
\end{align}
where \eqref{eq:state_duraton} indicates that $t_\mathrm{TB}$ cannot exceed the average duration of the state. Following this, we can set the block length $n(B,t_{\mathrm{TB}})=B t_{\mathrm{TB}}$. We follow equal duration channel partitioning where the thresholds are uniquely specified to provide equal state duration based on  $f_\mathrm{D}$ and $t_\mathrm{TB}$ values  \cite{zhang1999finite}. 

With crossover probabilities modeled by AWGN, the symbol level SNR when the channel gain is $|h(t)|=x$, is given by $\gamma(x,t)=\frac{P x^2}{B N_0}$. The normalized SNR at the $\ell^{th}$ channel state, denoted by $\Gamma_{\ell}$ for $L=1,\cdots,L$, is then given by \cite{sahin2019delay}:
\begin{align} 
\nonumber\Gamma_{\ell}&= \frac {\int_{\eta_{\ell}}^{\eta_{\ell+1}} { \gamma(x,t) 2 x e^{-x^{2}} dx }}{q_{\ell}( \eta_\ell,\eta_{\ell+1})}
\\&=\frac{P}{BN_0} \frac {e^{-\eta_\ell^2}(\eta_\ell^2+1)-e^{-\eta_{\ell+1}^2}(\eta_{\ell+1}^2+1)}{e^{-\eta_\ell^2}-e^{-\eta_{\ell+1}^2}}.
\label{eq:SNR_soft2}
\end{align}

\subsection{Standard Hybrid Automatic Repeat Request (HARQ)}
As shown in Fig. \ref{fig:M_N-HARQ}(b) in IR-HARQ, the transmitter encodes the packet using a $(k,n)$ channel code and sends the initial $n$ symbols in the first transmission. If the receiver correctly decodes the message, it sends an acknowledgment (ACK) back to the transmitter. Upon receiving ACK, the transmitter discards the remaining parity symbols of the message and moves to the next packet, and sends $n$ initial symbols of the next codeword. If the receiver is unable to decode the message with $n$ symbols of the first transmission, it sends a NACK signal and requests to transmit additional $n_1=\tau_1n$ parity symbols. This continues until the transmitter receives an ACK or the maximum number of retransmissions is reached. 

In the case of CC-HARQ, the transmitter encodes the packet using a $(k,n)$ channel code and repeats it in the next time slot if a NACK is received. Note that in the conventional HARQ, each retransmission occupies a separate time slot after the first transmission. The receiver combines the retransmitted packet with the first packet using maximum ratio combining (MRC) or incremental redundancy combining (IRC) in CC-HARQ or IR-HARQ, respectively. Since each packet is transmitted independently in separated time slots, we refer to the standard HARQ scheme as orthogonal HARQ (O-HARQ). 
Using normal approximation \cite{polyanskiy2010channel}, the error rate in the FBL can be characterized for single use of channel with a specific SNR. For CC-HARQ, bound in \cite{polyanskiy2010channel} can be directly used with accumulated SNR after MRC.
With an initial transmission of a packet of length $n_0=n$ symbols followed by $(m-1)$ retransmissions, the error rate in CC-HARQ  can be written as
\begin{align}
    {\epsilon}_\mathrm{cc}\left([\gamma_i]_0^{m-1}\right)\approx Q\left( \frac{n \log_2(1+\sum_{i=0}^{m-1} \gamma_i)-k+\log_2(n)}{n \sqrt{   V(\sum_{i=0}^{m-1} \gamma_i)}}\right),
    \label{eq:CC-HARQ_AW}
 \end{align}
The bound developed in \cite{erseghe2016coding}, for parallel AWGN channels can be used to calculate packet error rate for IR-HARQ as follows\footnote{This bound was also used in \cite{sahin2019delay} to analyze the performance of IR-HARQ schemes under the FBL assumption}:
\begin{align}
\nonumber{\epsilon}&_\mathrm{ir}\left([\gamma_i]_0^{m-1}, [n_i]_0^{m-1}\right)\approx \\
    & Q\left( \frac{\sum_{i=0}^{m-1} n_i \log_2(1+\gamma_i)-k+\log_2(\sum_{i=0}^{m-1} n_i)}{\sqrt{\sum_{i=0}^{m-1}  n_i V(\gamma_i)}}\right),
    \label{eq:IR-HARQ_AW}
 \end{align}
where $[x]_1^m=[x_1,\cdots,x_m]$, $V(\gamma_i)= \left(1-(1+\gamma_i)^{-2}\right) \log_2^2(e)$ is the channel dispersion, $\gamma_i\in[\Gamma_{\ell}]_{1}^{L}$ is the SNR at the $i$-th round transmission, and $Q(.)$ is the standard $Q$-function.

\section{Non-orthogonal HARQ}
\label{sec:NOMA_MODEL}
\begin{figure}[t]
\centering
\includegraphics[scale=0.65]{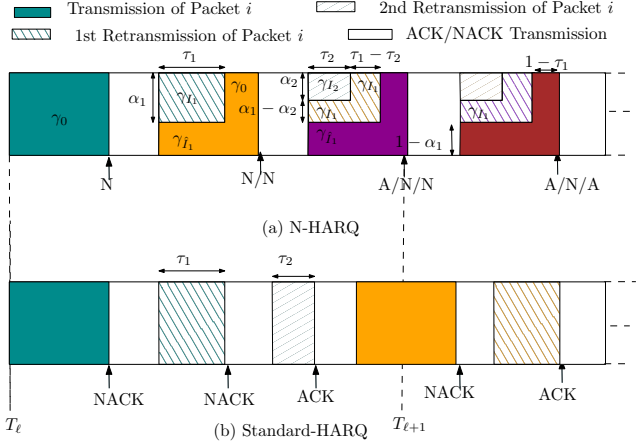}
\caption{The packet transmission in (a)  non-orthogonal HARQ and (b) standards IR-HARQ with $m=3$.}
\label{fig:M_N-HARQ}
\end{figure}
In conventional HARQ, when radio resources, such as bandwidth, are limited, the retransmission occupies an extra time slot. This leads to queuing and packet delivery delays that are deleterious to delay-sensitive communications. We propose a novel HARQ scheme that is based on non-orthogonal retransmissions and is suitable for delay-sensitive applications. In the proposed non-orthogonal HARQ (N-HARQ), the retransmission request is served together with the next packet by partially sharing the same time slot through superposition of signals non-orthogonally. In particular, for N-HARQ with maximum $(m-1)$ retransmissions, the transmitter superimposes maximum $m$ packets over consecutive $m$ time slots with power-splitting and time-sharing ratios $[\alpha]_1^{m-1}$ and $[\tau]_1^{m-1}$, respectively. In particular, the $i$-th retransmission of a packet will be conducted over a fraction $\tau_i$ of the time slot with power-splitting ratio $\alpha_i$. The SINRs of retransmitting signals vary due to non-orthogonal sharing of other retransmitting packets and the new packet over a single time slot. 

Fig. \ref{fig:M_N-HARQ} (a) represents the packet transmission using N-HARQ for maximum 2 retransmissions i.e., $m=3$. As can be seen, a time slot is partially shared between the new packet and the retransmissions of the previous packets.  In the example provided in Fig.\ref{fig:M_N-HARQ} (a), in the first slot, the new packet is not sharing the slot with any retransmission therefore, the packet is received with SNR $\gamma_0=|h|^2P/N_0$ at the receiver. Since this packet is not decoded successfully, in the next time slot, the 1st retransmission is partially superimposed with the new packet in the $\tau_1$ fraction of the time slot with power-splitting ratios $\alpha_1$. The retransmission and new packets are received with SINR $\frac{\alpha_1\gamma_0}{(1-\alpha_1)\gamma_0+1}$ and  $\frac{(1-\alpha_1)\gamma_0}{\alpha_1\gamma_0+1}$ over the $\tau_1$ fraction of the time slot, respectively, while the new packet is received with SINR $\gamma_0$ over the rest of the time slot. As both packets failed the decoding, second and first retransmissions of previous packets will be partially superimposed with the new packet over $\tau_2$ and $\tau_1$ fraction of the time slot with power-splitting ratios $\alpha_2$ and $\alpha_1$. The SINR of the 2nd retransmission, the 1st retransmission of previous packets, and the new packet in the $\tau_2$ fraction of the time slot are $\frac{\alpha_2\gamma_0}{(1-\alpha_2)\gamma_0+1}$, $\frac{(\alpha_1-\alpha_2)\gamma_0}{(1-\alpha_1+\alpha_2)\gamma_0+1}$, and $\frac{(1-\alpha_1)\gamma_0}{\alpha_1\gamma_0+1}$, respectively. Accordingly, the SINR of the 1st retransmission of the previous packet and the new packet over $\tau_1-\tau_2$ fraction of the time slot is given by $\frac{\alpha_1\gamma_0}{(1-\alpha_1)\gamma_0+1}$ and $\frac{(1-\alpha_1)\gamma_0}{\alpha_1\gamma_0+1}$, respectively. The SINR of the new packet over $1-\tau_1$ fraction of the time slot will be $\gamma_0$. Note that we assume $N_0=1$. 

The receiver combines each retransmission with previous packets to increase the decoding reliability until maximum $m$ rounds are met on which packet failure is declared. In case  packet is decoded successfully, its interference can be removed from other packets to improve their decoding probability. In general, the decoding order is from the packet with the highest overall SINR to that with the lowest overall SINR; however,  we always decode the packet that has arrived earlier in order to follow the strict packet delivery deadline of each packet. It is important to note that we assume that the receiver sends instantaneous feedback carrying $m$ bit information, which indicates the status of the decoding of each of the $m$ superimposed packets.

\section{Reliability, Throughput, and Delay Analysis of N-HARQ}
\label{sec:reliability_delay_analysis}
In this section, we present a detailed analysis of N-HARQ in the AWGN channel. We then show how this can be extended to the fading channel.
\subsection{N-HARQ with single retransmission over AWGN channel}
In N-HARQ with IR and maximum 1 retransmission, the SINR for the retransmitted packet and the new packet sharing the $\tau_1$ fraction of the time slot can be written as  $\gamma_{1}=\frac{\alpha_1 \gamma_0}{1+(1-\alpha_1) \gamma_0}$ and  $\gamma_{e}=\frac{(1-\alpha_1) \gamma_0 }{1+\alpha_1 \gamma_0}$, respectively, where $0\le\alpha_1\le 1$ is the power-splitting ratio during the retransmission and $\gamma_0=|h|^2P/N_0$. Due to partial retransmissions, the remaining $1-\tau_1$ duration of the codeword is sent without interference with SNR  $\gamma_0$. When no retransmission is requested for the  packet, the SNR for the newly sent packet is $\gamma_0$. The IR receiver combines the  $n_1=\tau_1 n$ retransmission parity symbols with the first transmission symbols and then decodes the longer codeword to achieve higher reliability. In IR-HARQ, SIC is used to remove the successfully decoded symbols from the  overlapping new arriving packet in the partial sharing interval of a time slot. In CC, the power-splitting is enabled for the entire new time slot, i.e., $\tau_1=1$, and the MRC receiver is used to improve the effective SINR. Therefore, in CC after successfully decoding a packet, SIC is utilized to remove its interference from the new packet. 
\begin{figure}[t]
\centering
\includegraphics[width=0.7\columnwidth]{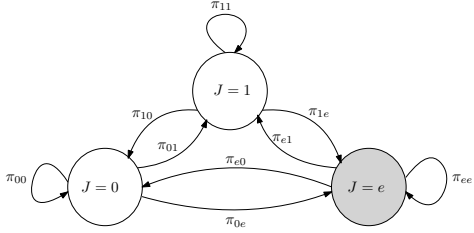}
\caption{The Markov Model for N-HARQ with single retransmission in AWGN channel.}
\label{fig:MarkovModel1}
\end{figure}

We use the Markov model to analyze the dynamic of the proposed N-HARQ in the FBL regime. Fig. \ref{fig:MarkovModel1} shows the Markov model of the proposed N-HARQ with single retransmission, where each state, denoted by $J$, represents the number of retransmissions. That is at state $J=0$, the packet has been successfully decoded with only 1 transmission , and therefore, the time slot is 
fully used to transmit
the newly generated packet. 
At state $J=1$, the packet has been successfully decoded with 1 retransmission. At state $J=e$, the packet has failed the decoding even after 1 retransmission. Let $\Pi=[\pi_{ij}]$ denote the state transition matrix for N-HARQ with maximum 1 retransmission,  i.e, $m=2$, where $\pi_{ij}$ is the probability of transitioning from state $J=i$ to state $J=j$, for $i,j\in\{0,~1,~e\}$. The following lemma characterized the state transition probabilities.

\begin{lemma}
\label{lem:AWGN_pij_m2}
In N-HARQ with single retransmission ($m=2$) and incremental redundancy, the state transition probabilities are given by:
\[\pi_{ij}=\left\{\begin{array}{ll}
1-{\epsilon}_\mathrm{ir}\left([\gamma_i,\gamma_0],[\tau_1,1-\tau_1]\right);&j=0,\\
{\epsilon}_\mathrm{ir}\left([\gamma_i,\gamma_0,\gamma_I],[\tau_1,1-\tau_1,\tau_1]\right);&j=e,\\
1-\pi_{i0}-\pi_{ie};&j=1,\\
\end{array}\right.\]
where $\gamma_0=\frac{P|h|^2}{N_0}$, $\gamma_1=(1-\alpha_1)\gamma_0$, $\gamma_e=\frac{(1-\alpha_1)\gamma_0}{1+\alpha_1\gamma_0}$, and $\gamma_I=\frac{\alpha_1\gamma_0}{1+(1-\alpha_1)\gamma_0}$. 
\end{lemma}

\begin{IEEEproof}
The system transits from state $J=i$ to state $J=0$ for $i\in\{0,1,e\}$ if the decoding is successful, with probability $1-\epsilon_{\mathrm{ir}}([\gamma_i,  \gamma_0], [{\tau_1}, {1-\tau_1}])$, by receiving only one packet with SINR $\gamma_i$ over $\tau_1$ fraction  and SNR $\gamma_0$ over remaining $1-\tau_1$ fraction of the time slot, where $\gamma_0=|h|^2P/N_0$ as there is no interference at state $J=0$, $\gamma_1=(1-\alpha_1)\gamma_0$ since at state $J=1$ the previous packet has been decoded successfully and the interference was removed, and $\gamma_e=\frac{(1-\alpha_1)\gamma_0}{1+\alpha_1\gamma_0}$ as at state $J=e$ the previous packet was not decoded successfully. The system transits to state $J=e$ if the decoding failed after two transmissions, which happened with probability ${\epsilon}_\mathrm{ir}\left([\gamma_i,\gamma_0,\gamma_I],[\tau_1,1-\tau_1,\tau_1]\right)$. The system transits to state $J=1$ if the decoding succeeds with exactly one retransmission. That is 
 \begin{align}
     \nonumber\pi_{i1}&=\mathrm{Prob}\{\mathcal{A}_1(i) \wedge \mathcal{A}^{c}_1(i)\}\\
     \nonumber&=\mathrm{Prob}\{\mathcal{A}_1(i)\}-\mathrm{Prob}\{ \mathcal{A}_1(i)\wedge \mathcal{A}_2(i)\}\\
     \nonumber &\overset{(a)}{\approx} \mathrm{Prob}\{\mathcal{A}_1(i)\}-\mathrm{Prob}\{\mathcal{A}_2(i)\}\\
     &=\epsilon_\mathrm{ir}([\gamma_i,\gamma_0],[\tau_1, {1-\tau_1}])-\epsilon_\mathrm{ir}([\gamma_i,\gamma_0, \gamma_I],[\tau_1, 1-\tau_1,\tau]),
 \end{align}
where $\mathcal{A}_0(i)$ and $\mathcal{A}_1(i)$ denote the events that the decoding failed without or with single retransmission at state $J=i$, respectively. Step $(a)$ follows from the fact that if the packet is failed after a retransmission, it almost certainly fails without retransmission.
\end{IEEEproof}
\begin{remark}
Lemma \ref{lem:AWGN_pij_m2} reduces to \cite[Lemma 1]{nadeem2020non} when $\tau_1=1$. 
\end{remark}
\begin{remark}
It is easy to verify that for N-HARQ with single retransmission ($m=2$) and Chase combining when $\tau_1=1$, the state transition probabilities are given by:
\[\pi_{ij}=\left\{\begin{array}{ll}
1-{\epsilon}_\mathrm{cc}\left([\gamma_i]\right);&j=0,\\
{\epsilon}_\mathrm{cc}\left([\gamma_i,\gamma_I]\right);&j=e,\\
1-\pi_{i0}-\pi_{ie};&j=1,\\
\end{array}\right.\]
where $\gamma_0$, $\gamma_1$, $\gamma_e$ and $\gamma_I$ are defined in Lemma \ref{lem:AWGN_pij_m2}.
\end{remark}
\begin{remark}
Let $P_\text{stat}=[p_0, p_1, p_e]$ denotes the stationary distribution corresponding to state transition matrix $\Pi$. The PER of the N-HARQ with single retransmission, denoted by $\zeta$ is given by:
\begin{align}
       \zeta=p_e=\left(1+\frac{\pi_{01}\pi_{e0}+\pi_{e1}\left(2+\pi_{10}-\pi_{11}-\pi_{00}\right)}{(\pi_{00}-1)(\pi_{11}-1)-\pi_{01}\pi_{10} }\right)^{-1}.
\label{eq:packet_error_rate}
\end{align}
\end{remark}
This follows directly from the fact that the
stationary distribution of a Markov chain can be characterized by the eigenvector of matrix $\Pi^\mathrm{T}$, corresponds to eigenvalue $1$, and the packet error probability is simply the stationary probability of being at state $J = e$. By solving a set of linear equations  $(\Pi-I_3)P^{\mathrm{T}}_\mathrm{stat}=\mathbf{0}$, where $I_3$ is a 3$\times$3 identity matrix, (\ref{eq:packet_error_rate}) can be easily derived.

\begin{remark}
\label{lemma-th-noma}
The throughput of N-HARQ, denoted by $\eta$, is given by \cite{nadeem2020non}:
\begin{align}
 \eta=\frac{k(1-\zeta)}{n}.
\label{eq:thrpt_NOMA_HARQ}
\end{align}
\end{remark}

As in N-HARQ there is no queuing; we send at most $N+1$ packets when $N$ packets are scheduled for the transmission. Then the packet delivery delay profile for N-HARQ with single retransmission can be characterized as follows:
\begin{align}
    D^{(N)}_\text{N}[d]=p_0\delta[d-N]+(1-p_0)\delta[d-N-1],
    \label{eq:pdf_noma}
\end{align}
 where $\delta[d]$ is the discrete Dirac delta function and $p_0$ is obtained from the stationary distribution. 
 
\subsection{N-HARQ with two retransmissions over the AWGN channel}
Now we provide detailed reliability and delay analysis of the  N-HARQ in the AWGN channel when two retransmissions are enabled.  When $m=3$, the Markov model has 4 states where $J=2$ represents that the packet is successfully decoded at the receiver with 2 retransmissions. The system transits to state $J=e$ if the decoding is failed after two retransmissions. Note that at state $J=2$ and $J=e$ the state is partially occupied with the new arriving packet for  $\tau_2$ duration while the remaining $1-\tau_2$ is free from interference. At state $J=2$ and $J=e$ the time slot is occupied with 3 packets. The following lemma characterizes the state transition matrix.
\begin{lemma}
\label{lem:Pim3awgn}
Let  $\Pi=[\pi_{ij}]$ denotes the state transition matrix for N-HARQ with maximum 2 retransmissions, i.e $m=3$, where $\pi_{ij}$  denotes the probability of transiting from State $J=i$ to State $J=j$.  Then for $i\in\{0,1\}$ $j\in\{0, 1, 2, e\}$, $\pi_{ij}$ is given by:
\[\pi_{ij}=\left\{\begin{array}{lll}
1-\mathrm{Prob}\{\mathcal{A}_0\}; & j=0,\\
    \mathrm{Prob}\{\mathcal{A}_0\}-\mathrm{Prob}\{\mathcal{A}_1\}; & j=1,\\ 
    \mathrm{Prob}\{\mathcal{A}_1\}-\mathrm{Prob}\{\mathcal{A}_2\}; & j=2,\\
 \mathrm{Prob}\{\mathcal{A}_2\}; & j=e,\\ 
\end{array}\right.\]
where  $\mathcal{A}_0$, $\mathcal{A}_1$ and $\mathcal{A}_2$ represents events that the decoding failed without retransmission, with only one retransmission, and with two retransmissions, respectively.  We further have $\mathrm{Prob}\{\mathcal{A}_0\}= \epsilon_\mathrm{ir}([\gamma_{i},\gamma_0], [\tau_1, {1-\tau_1}])$ ,  $\mathrm{Prob}\{\mathcal{A}_1\}= \epsilon_\mathrm{ir}([\gamma_{i},\gamma_0, \gamma_{I_1}], [\tau_1, {1-\tau_1}, \tau_1])$ and  $\mathrm{Prob}\{\mathcal{A}_2\}=\epsilon_\mathrm{ir}([\gamma_{i},\gamma_0,$  $\gamma_{I_1},\gamma_{I_2}], [\tau_1, {1-\tau_1}, \tau_1, \tau_2])$, where $\gamma_0=P/N_0$, $\gamma_1=(1-\alpha_1)P+N_0$, $\gamma_{I_1}=\frac{\alpha_1P}{(1-\alpha_1)P+N_0}$, and $\gamma_{I_2}=\frac{\alpha_2P}{(1-\alpha_2)P+N_0}$ and $i\in\{0,1\}$. Similarly for $i\in\{2,e\}$ and $j\in\{0, 1, 2, e\}$, $\pi_{ij}$ is given by:
\[\pi_{ij}=\left\{\begin{array}{lll}
\sum_{k}p_k(1-\mathrm{Prob}\{\mathcal{B}_0(k)\}); & j=0,\\
   \sum_{k}p_k( \mathrm{Prob}\{\mathcal{B}_0(k)\}-\mathrm{Prob}\{\mathcal{B}_1(k)\}); & j=1,\\ 
  \sum_{k}p_k(\mathrm{Prob}\{\mathcal{B}_1(k)\}-\mathrm{Prob}\{\mathcal{B}_2(k)\}; & j=2,\\
 \sum_{k}p_k(\mathrm{Prob}\{\mathcal{B}_2(k)\}); & j=e,\\ 
\end{array}\right.\]
where $\mathcal{B}_0(k)$, $\mathcal{B}_1(k)$ and $\mathcal{B}_2(k)$ represent events that the decoding failed with $0$, $1$, and $2$ retransmissions, respectively, when another packet in state $k\in\{0,1,2,e\}$ is present in-between the state transition from $J=i$ to $J=j$, and $p_k$ is the stationary probability of state $J=k$.  
\end{lemma}
\begin{IEEEproof}
Please refer to Appendix \ref{app:prooflemmam3} for the detailed proof.
\end{IEEEproof}
Let $P_{\mathrm{stat}}=[p_0, p_1, p_2, p_e]$ denotes the stationary distribution corresponding to the state transition matrix $\Pi$ when $m=3$. Then, the stationary probabilities can be formulated as $(\Pi-I_4)P^{\mathrm{T}}_\mathrm{stat}=\mathbf{0}$, where $I_4$ is a 4$\times$4 identity matrix. This however results in a system of nonlinear equations due to conditional probabilities as shown in the previous lemma. In order to find stationary probabilities, we use iterative methods to solve the system of non-linear equations. We use state of the art sequential quadratic programming (SQP) technique to solve the non-linearly constrained problem  \cite{hock1983comparative}. The throughput can then be found by using (\ref{eq:thrpt_NOMA_HARQ}).

As in N-HARQ, there is no queuing; we send at most $N+2$ packets when $N$ packets are scheduled for the transmission with maximum of $2$ retransmission. Then the packet delivery delay profile for N-HARQ with maximum 2 retransmissions can be characterized as follows
\begin{align}
    \nonumber D^{(N)}_\text{N}[d]&=p_0\delta[d-N]+p_1\delta[d-N-\tau_1]\\
    &+(1-p_0-p_1)\delta[d-N-(1+\tau_2)],
    % \label{eq:pdf_noma}
\end{align}
where $p_0$ and $p_1$ are obtained from the stationary distribution.  
\subsection{Optimization of retransmission coefficients}
In N-HARQ, the power-splitting and time-sharing parameters can be optimized in order to maximize the throughput without delaying packets due to queuing. In AWGN, the PER minimization for target throughput $\eta_0$, given $k$, $n$, the maximum number of transmissions $m$, and transmit power $P$ when $\alpha_0=\tau_0=1$, can be summarized as follows:
\begin{align}
\label{eq:opt}
&\min_{[\tau]_1^{m-1},[\alpha]_1^{m-1}}  \zeta\\
\nonumber \textrm{subject~to:}~\\
& \mathrm{C_1:}~ \eta\ge \eta_0, \nonumber\\
&\mathrm{C_2:}~0\le \alpha_i\le 1,~ \alpha_{i-1}\leq \alpha_i,\quad  \forall i \in [j]_1^{m-1},\nonumber\\ 
&\mathrm{C_3:}~  0\le \tau_i\le 1,~ \tau_{i-1}\leq \tau_i, \quad  ~~\forall i \in [j]_1^{m-1}, \nonumber
\end{align}
where $\mathrm{C_1}$ is imposed to guarantee the minimum throughput $\eta_0$ and $\mathrm{C_2}$ and $\mathrm{C_3}$ are the assumptions for the power-splitting and time-sharing parameters, respectively, in each retransmission.

\subsection{Analysis of N-HARQ over FSMC}
\begin{figure}[t]
\centering
\includegraphics[width=0.8\columnwidth]{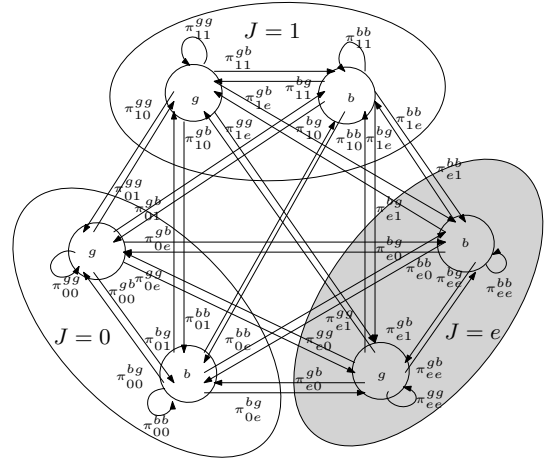}
\caption{The Markov Model for the proposed N-HARQ with maximum 1 retransmission  and $L=2$ channel fading states.}
\label{fig:MarkovModel}
\end{figure}

The analysis presented for N-HARQ in the AWGN channel can be easily extended to the  Rayleigh channel. In particular, when the system is at state $J=j$ for $j\in\{0,1,\cdots,m-1,e\}$, the channel can be at one of the states $S_{\ell}$ for $\ell\in[i]_1^L$. Therefore, the Markov model can be expanded to include both the retransmission and channel states. Fig. \ref{fig:MarkovModel} shows the MM for N-HARQ with single retransmission when the channel is represented by $L=2$ states, i.e., Good and Bad. 

Let $\Pi_\mathrm{Ray}=[\mathbf{\pi}_{ij}({\ell,k})]$ denotes the state transition probability matrix when the system transit from state $J=i$ to state $J=j$ while the fading state transit from $S_\ell$ to $S_k$ for  $i,j\in\{0,1,e\}$ and $\ell,k \in[i]_1^L$. The following lemma characterizes $\Pi_\mathrm{Ray}$.  

\begin{lemma}
In N-HARQ with single retransmission, the state transition probabilities $\pi_{i,j}({k,\ell)}$ for $i,j \in \{0,1,e\}$, $\ell,k\in \{0,1,\cdots, L\}$, with single partial retransmission, are defined as follows:
\[\pi_{ij}({\ell,k})=\left\{\begin{array}{ll}
P_{\ell,k}\left(1-\epsilon_\mathrm{ir}\left([\gamma_i(\ell),\gamma_0(\ell)],[\tau_1, {1-\tau_1}]\right)\right);&j=0,\\
P_{\ell,k}\epsilon_\mathrm{ir}([\gamma_i(\ell),\gamma_0(\ell),\gamma_I(k)],[{\tau_1}, {1-\tau_1},{\tau_1}]);&j=e,\\
P_{\ell,k}-\pi_{i,0}({\ell,k})-\pi_{i,e}({\ell,k});&j=1,\\
\end{array}\right.\]
where $P_{\ell,k}$ is the probability of transiting from the fading state $S_{\ell}$ to $S_k$, which is given by (\ref{eq:fadingtransitions}) for $|\ell-k|\le1$ and equals to $0$ otherwise,  $\gamma_0(\ell)=\Gamma_\ell$, $\gamma_1(\ell)=(1-\alpha)\Gamma_\ell$,
$\gamma_{I}(\ell)=\frac{\alpha \Gamma_\ell}{1+(1-\alpha) \Gamma_\ell}$ and
$\gamma_e(\ell)=\frac{(1-\alpha)\Gamma_\ell}{1+ \alpha \Gamma_\ell}$, where $\Gamma_\ell$ is the SNR over the $\ell$-th fading state, which is given by \eqref{eq:SNR_soft2}. 
\end{lemma}
\begin{IEEEproof}
The time block duration is set according to \eqref{eq:state_duraton} so that the average fading state duration is equal to the packet duration. Therefore, the packet state transition is independent to the channel fading state transition. The lemma then directly follows from the proof of Lemma \ref{lem:AWGN_pij_m2}.
\end{IEEEproof}

Let  vector ${P}_\mathrm{stat}=[p_0(\ell),p_1(\ell), \cdots, p_e(\ell) ]_{1\times3L}$, for $\ell \in \mathcal{L}$ denote vector containing stationary distribution of each packet state when the fading state is $S_\ell$. Then the stationary distribution can be obtained by solving  a set of linear equations  $(\Pi_\mathrm{Ray}-I_{3L})P_\mathrm{stat}^{\mathrm{T}}=\mathbf{0}$, where $I_{3L}$ is $3L\times3L$ identity matrix. The stationary probability of being at state $J = i$ is  $p_i=\sum_{\ell\in \mathcal{L}} p_i(\ell)$, where $i=\{0, 1, e\}$. Then the PER of the  proposed scheme is given by $\zeta=\sum_{\ell\in \mathcal{L}} p_e(\ell)$. The throughput and delay profile can be characterized using the stationary distribution by  \eqref{eq:thrpt_NOMA_HARQ} and \eqref{eq:pdf_noma}, respectively. An optimization problem similar to (\ref{eq:opt}) can also be defined to minimize PER in the Rayleigh fading channel for the desired target throughput.

\subsection{Reliabity and delay analysis of O-HARQ}
\label{APP:Baseline_standard_HARQ_analysis}
In standard HARQ, retransmissions are carried out irrespective of the new arriving messages. As shown in Fig. \ref{fig:M_N-HARQ}(b), the retransmissions are conducted in separate time slots with full power, and packets are transmitted without interfering each other. Therefore, all packets, including their retransmissions, have the same SNR equals to $\gamma_0$. Let $p_i$ denotes the probability that the decoding of a packet is successful after exactly $i$ retransmissions. In IR-HARQ, $p_i$ for $i\in[j]_{1}^{m}$ is given by:
\begin{align}
    p_{i}\approx\epsilon_{\mathrm{ir}}\left(\gamma_0,\sum_{j=0}^{i-1}\tau_j\right)-\epsilon_{\mathrm{ir}}\left(\gamma_0,\sum_{j=0}^{i}\tau_j\right),
\end{align}
where $\tau_0=1$, $p_0=1-\epsilon_{\mathrm{ir}}(\gamma_0,\tau_0)$ and  $p_e=\epsilon_{\mathrm{ir}}\left(\gamma_0,\sum_{j=0}^{m}\tau_j\right)$ is the PER. For CC-HARQ, $p_i$ is given by:
\begin{align}
    p_{i}\approx\epsilon_{\mathrm{cc}}\left((i-1)\gamma_0,\tau_0\right)-\epsilon_{\mathrm{cc}}\left(i\gamma_0,\tau_0\right), 
\end{align}
where $\tau_0=1$, $p_0=1-\epsilon_{\mathrm{cc}}(\gamma_0,\tau_0)$ and  $p_e=\epsilon_{\mathrm{cc}}\left(m\gamma_0,\tau_0\right)$ is the PER. The throughput of standard HARQ with $m$ retransmissions is then given by:
\begin{equation}
  \eta= \frac{k}{n} \times \frac{1-p_e}{ \sum_{i=0}^{m}p_{i}\sum_{j=0}^{i}\tau_{j} +   p_{e}\sum_{i=0}^{m} \tau_{i} },
  \label{eq:throughput_AWGN}
\end{equation}
and the delay distribution for a given packet with standard HARQ is given by:
\begin{align}
    D[d]=p_0\delta[d-1]&+\sum_{i=1}^{m}p_{i}\delta\left[d-\sum_{j=0}^i\tau_{j}\right]+p_e\delta\left[d-\sum_{i=0}^{m}\tau_{i}\right],
    \label{eq:Delay_distrm}
\end{align}
where $\delta[d]$ is the discrete Dirac delta function, and for CC-HARQ we have $\tau_i=\tau_0$ for $i\in[j]_1^{m}$. It is then easy to show that the overall delay distribution for delivering $N$ packets (either successful or unsuccessful) with standard IR-HARQ can be found as follows:
\begin{align}
D^{(N)}_\text{O}[d]=\bigotimes_{i=1}^ND[d],
\end{align}
where $\otimes$ is the convolution operand. For example, at $m=1$, the delay distribution of delivering $N$ packets is given by \cite{nadeem2020non}:
\begin{align}
    D^{(N)}_\text{O}[d]=\sum_{i=0}^{N}\dbinom{N}{i}(1-\epsilon(\gamma_0,1))^{i}&\epsilon(\gamma_0,1)^{N-i}\delta[d-(1+\tau_1)N+i],
    \label{eq:Qdelay_N1}
\end{align} 

In Rayleigh fading channel, The success and fail probabilities when $m=1$ for standard HARQ can be calculated directly using marginal  probabilities of each fading state and its transitioning probabilities as follows:
\begin{align}
    p_0=&\sum_{\ell\in\mathcal{L}}q_{\ell}\left(1-\epsilon\left(\Gamma_\ell,\tau_0\right)\right)\\ \nonumber
     p_1=&\sum_{\ell\in\mathcal{L}}\sum_{k\in\mathcal{L}}q_{\ell}P_{\ell,k}(\epsilon(\Gamma_\ell,\tau_0)-\epsilon([\Gamma_\ell,\Gamma_k],[\tau_0,\tau_1]))\\ \nonumber
     p_e=&\sum_{\ell\in\mathcal{L}}\sum_{k\in\mathcal{L}}q_{\ell}P_{\ell,k}\epsilon([\Gamma_\ell,\Gamma_k],[\tau_0,\tau_1])
\end{align}
where $q_\ell$ is the marginal probability of the $\ell$-th fading state, which is given in \eqref{eq:marg_prob_states}, and $P_{\ell,k}$ is the transition probability between fading state $\ell$ and $k$ that is given in \eqref{eq:fadingtransitions}. The throughput and delay distribution can be calculated by using \eqref{eq:throughput_AWGN} and  \eqref{eq:Delay_distrm}, respectively.

\section{Numerical Results}
\label{Sec:Numarical_resutls}
In this section, we present simulation results to evaluate the reliability and delay performance of N-HARQ in comparison with standards O-HARQ. First, we present the performance comparison in the AWGN channel and then in the Rayleigh fading channel. 
In the simulations, we mainly focus on the IR-HARQ scheme for comprehensive analysis; however, CC-HARQ performance is also compared for some special cases. In the general setup, IR-HARQ performs better due to its coding gain than CC-HARQ, which improves the received SNR with retransmission. In CC-HARQ, the transmitter sends the entire packet as retransmission so, there is no additional synchronization signaling required. In contrast, IR-HARQ requires variable retransmission lengths with slight signaling overhead to track the varying redundancy.
\begin{figure*}[t]
    \centering
    \begin{subfigure}{.495\textwidth}
    \includegraphics[width=\columnwidth]{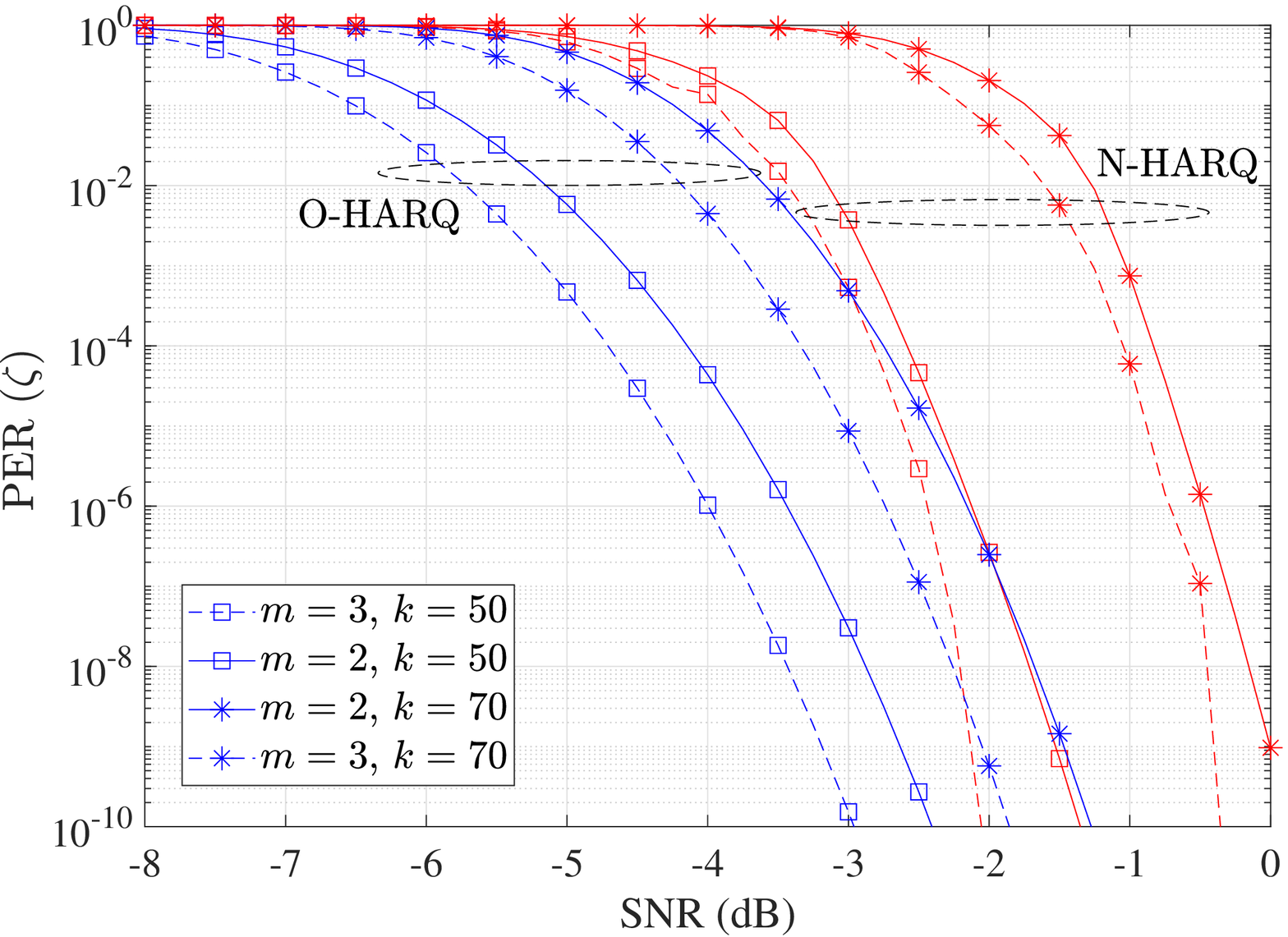}
    \caption{Packet error rate (PER)}
    \label{fig:PERm2m3} 
    \end{subfigure}
    \begin{subfigure}{.495\textwidth}
    \centering
    \includegraphics[width=\columnwidth]{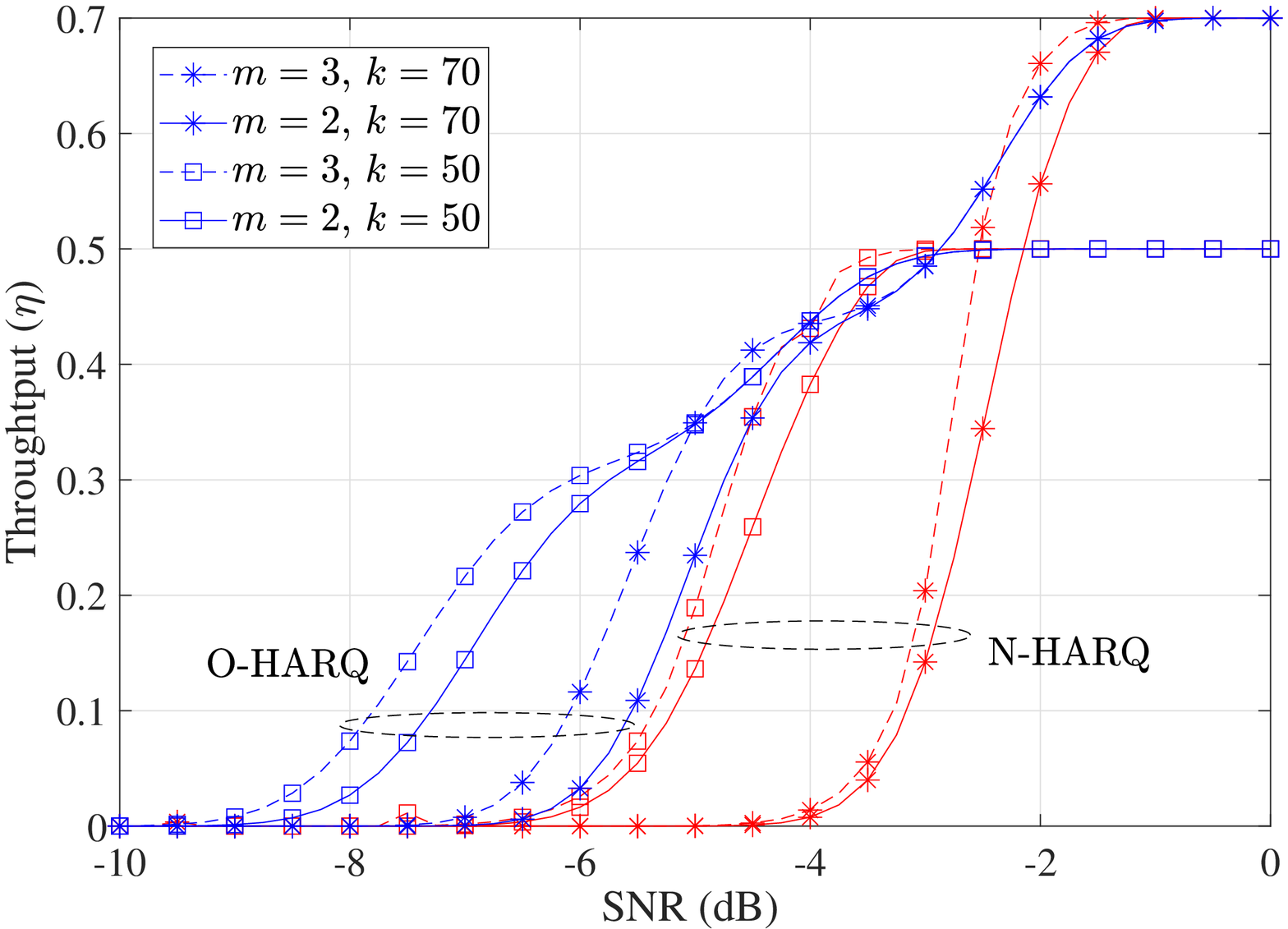}
    \caption{Throughput}
    \label{fig:thpm2m3}
    \end{subfigure}
    \caption{Comparison between N-HARQ-IR and O-HARQ in the AWGN channel when
    $n=100$. For $m=3$, $\alpha_1=0.7,\alpha_2=0.5, \tau_1=0.6, \tau_2=0.2$ and for $m=2$, $\alpha_1=0.7, \tau_1=0.6$. For O-HARQ, we use $\tau_1=0.6$ and $\tau_2=0.2$.}
    \vspace{-3ex}
\end{figure*}
\subsection{N-HARQ over the AWGN Channel}
In this section, we present numerical results for the proposed N-HARQ scheme with single and two retransmissions  in the AWGN channel. We compare the performance with the benchmark O-HARQ scheme. 
Fig. \ref{fig:PERm2m3} shows the PER versus SNR for the proposed N-HARQ in comparison with the standard O-HARQ at fixed block length $n = 100$. The code rate can be varied as $R = 0.5$ and $R=0.7$ by setting $k = 50$ and $k = 70$. In general, the 
PER performance of O-HARQ is better than N-HARQ for all SNRs,  rate $R$ and  retransmission number $m$. More specifically, for target PER $10^{-6}$, there is a SNR gap of $\sim2$ dB between N-HARQ and O-HARQ, when $m=2$ and $k=70$. This gap reduces to $\sim1.5$ dB by lowering the information rate, i.e., at $k=50$. We can observe a similar PER performance gap between O-HARQ and N-HARQ when two retransmissions are allowed, i.e., $m=3$ in Fig \ref{fig:PERm2m3}. This performance gap is because, in O-HARQ, retransmission are always carried out in separate time slots and with full power $P$, whereas in N-HARQ, retransmission packets share the time slot with new arriving packets, which leads to higher PER due to interference. 
Fig. \ref{fig:thpm2m3} shows the throughput of N-HARQ and O-HARQ at fixed block length $n=100$.  As can be seen, in the low-to-medium SNR regime, the throughput performance of N-HARQ is limited in comparison with O-HARQ. However, the performance gap reduces as the SNR increases. For example, when $k=70$, at SNR$=-4$dB, O-HARQ and N-HARQ achieve the same throughput. Note that the throughput gain with N-HARQ steadily improves from lower to higher SNR regimes, as in these regimes the interference signal can be decoded and removed with high probability. On the other hand, the throughout gain of O-HARQ saturates as retransmission always occupies a separate time slot. In contrast, in N-HARQ, the retransmission is served with the new packet transmission, which increases the throughput.

\begin{figure*}[t]
    \centering
    \begin{subfigure}{.495\textwidth}
    \includegraphics[width=\columnwidth]{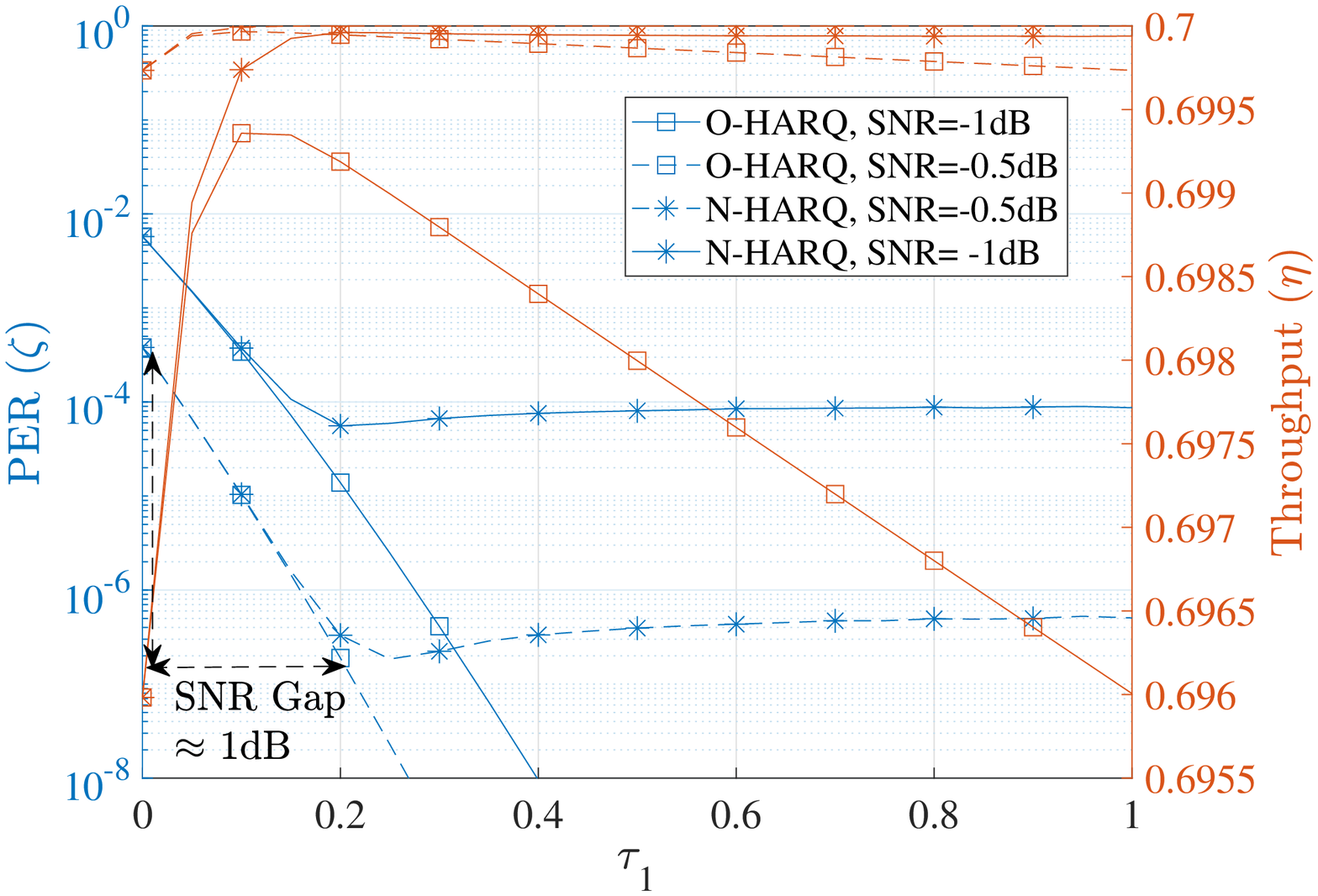}
    \caption{Reliability}
    \label{fig_delayawgn_a} 
    \end{subfigure}
    \begin{subfigure}{.495\textwidth}
    \centering
    \includegraphics[width=\columnwidth]{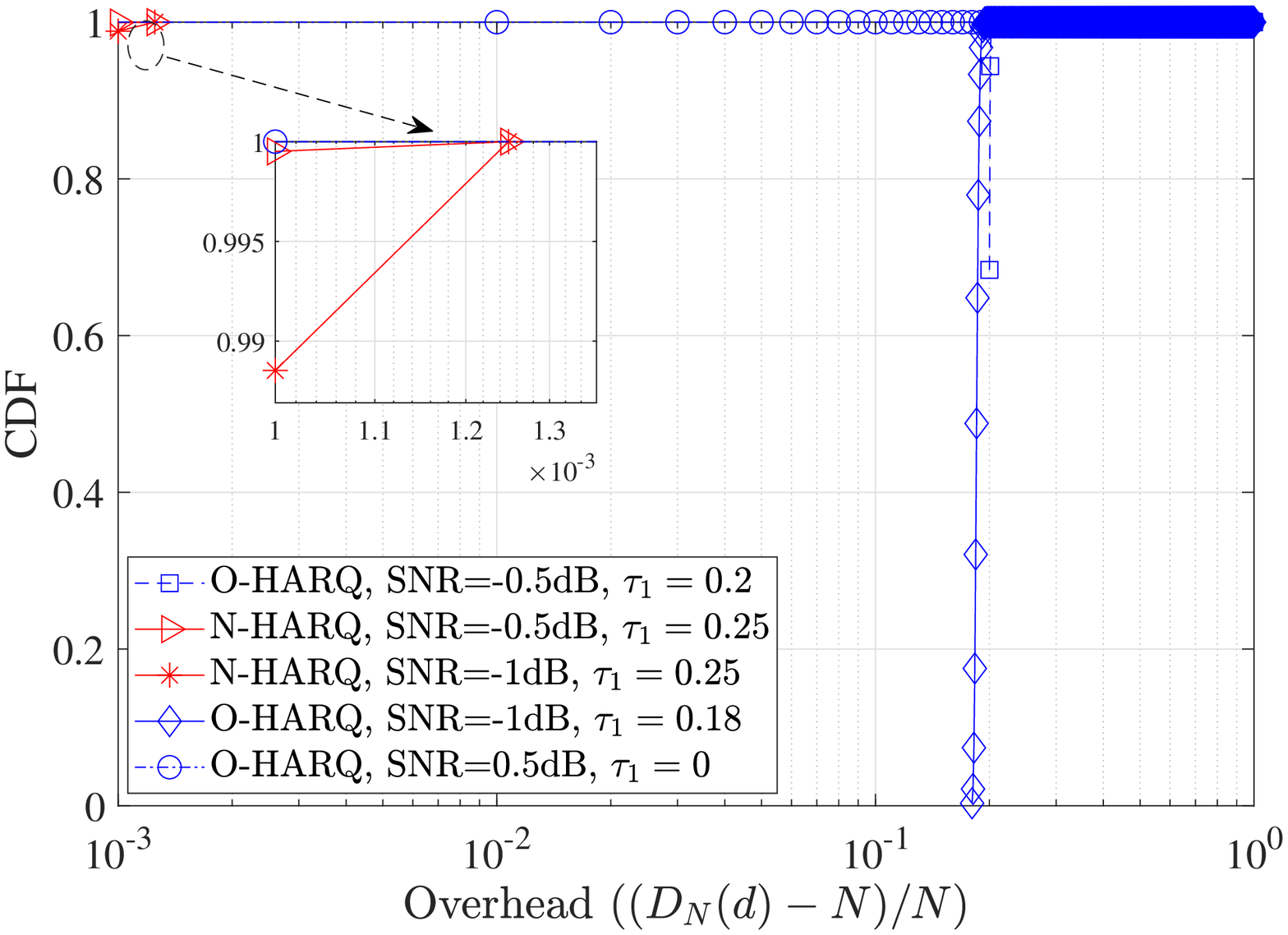}
    \caption{Latency}
    \label{fig_delayawgn_b}
    \end{subfigure}
    \caption{Delay performance comparison of  N-HARQ and  standard O-HARQ, at $n=100$, $k=70$, where we choose optimal $\alpha$ for N-HARQ for each $\tau_1$, and $N=1000$.}
    \label{fig_delayawgn}
    \vspace{-3ex}
\end{figure*} 

Fig. \ref{fig_delayawgn} shows the reliability and latency performance comparison between N-HARQ and O-HARQ while sending $N=1000$ packets. Firstly, Fig. \ref{fig_delayawgn_a} shows the PER and throughput performance variation with retransmission coefficient $\tau_1$. As can be seen in the figure, at fixed SNRs, in order to reduce PER, throughput performance of standard O-HARQ  degrades with more retransmissions, i.e., increasing $\tau_1$. Furthermore, Fig. \ref{fig_delayawgn_a} shows an optimal value of $\tau_1$ exists for PER minimization for a target  throughput,  and Fig. \ref{fig_delayawgn_b} shows the corresponding  delay  performance of O-HARQ and N-HARQ. As can be seen in Fig. \ref{fig_delayawgn}, O-HARQ suffers from large latency to achieve the target PER performance, while N-HARQ can achieve the same target PER performance with a significantly reduced latency. For example, at SNR=-1dB, N-HARQ can achieve PER$\approx$ $10^{-4}$ with throughput $\approx 0.7$ at $\tau=0.2$ without any delay overhead. However, O-HARQ achieves similar PER and throughput performance with $\tau_1=0.18$, which incurs $10\%$ delay overhead. In order for O-HARQ to achieve the same PER without additional delay, the transmission power needs to be increased. More specifically, for O-HARQ to achieve PER $\approx 10^{-7}$ with a latency performance similar to N-HARQ, the transmission power should be increased by 1dB. Otherwise, the PER performance degrades to  $\approx 10^{-3.5}$. This shows that N-HARQ has overall performance gain over the standard O-HARQ in terms of per-packet delay performance.

\begin{figure}[t]
    \centering
    \includegraphics[width=1\columnwidth]{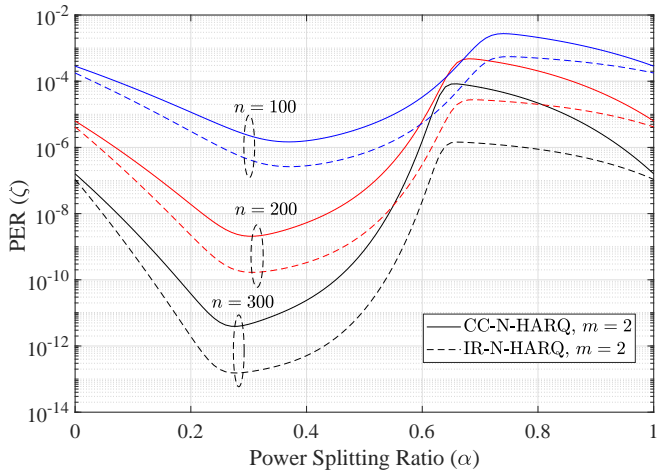}
    \caption{effect of $\alpha$ over PER performance of IR and CC non orthogonal HARQ with various codeword lengths keeping $k=n/2$ at SNR=-2dB. For IR-HARQ, we set $\tau=1$.}
    \label{fig:PER_V_N}
\end{figure}
In Fig. \ref{fig:PER_V_N}, we show the PER performance of N-HARQ with various packet lengths for Chase combining and incremental redundancy. As can be seen, for a given rate $k/n$, one can achieve lower PERs by increasing the block length $n$.  Note that due to fixed bandwidth, longer $n$ will utilize longer time blocks for its transmission. Overall, IR-N-HARQ outperforms CC-N-HARQ in terms of PER across all power-splitting ratios due to information combining receiver in comparison with the energy combining receiver in CC. Chase combining however, is much simpler in both the transmitter and receiver. It can be also observed in Fig. \ref{fig:PER_V_N} that for both CC and IR, there exists an optimal $\alpha$ where PER is minimized.

\begin{figure}[t]
    \centering
    \includegraphics[width=1\columnwidth]{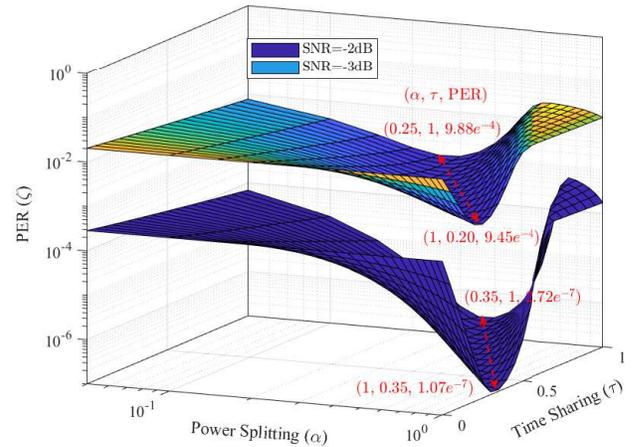}
    \caption{PER performance of IR-N-HARQ versus time-sharing and power-splitting ratios $\tau$ and $\alpha$, when $m=2$, $n=100$, and $k=50$.}
    \label{fig:PER_OPT_M1}
\end{figure}
Fig. \ref{fig:PER_OPT_M1} shows the PER performance variation with $\alpha$ and $\tau$ for IR-HARQ when $m=2$. As can be seen, PER reduces sharply by adjusting $\alpha$ and $\tau$. For example, in Fig. \ref{fig:PER_OPT_M1}, when SNR$=-2$dB at point $( \alpha=1,\tau=0.35)$, the PER of $10^{-7}$, which reduces with variation in $\tau$.  Note that adjusting $\tau$ increases the design complexity as it translates to codeword length variations.  However, as can be seen in the figure, at each $\tau$, there  exists an $\alpha$ that minimized the PER. For example at  $(\alpha=0.35,\tau=1)$, the PER of $2.7\times 10^{-7}$ is achieved. This shows that PER can be maintained by tuning power-splitting parameter $\alpha$ only while fixing $\tau$ to some constant in order to  gain implementation simplifications.

Note that the overall trend of the optimization points remain similar with change in SNRs, but the exact value of optimal point changes with SNRs.  This is an important design consideration for delay-sensitive applications, as slight SNR estimation error can limit the achievable performance. For example, the points on the boundaries are more susceptible to performance degradation due to SNR fluctuations than others.  

\begin{figure*}[t]
\begin{subfigure}{0.495\textwidth}
    \centering
    \includegraphics[width=\columnwidth]{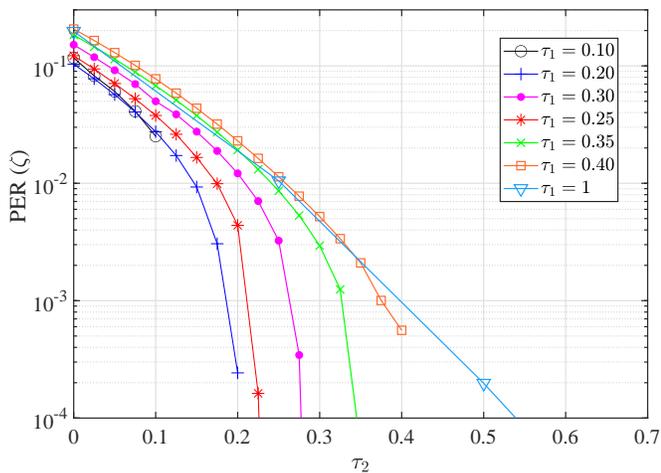}
    \caption{Packet Error Rate}
    \label{fig:PERa1a2}
\end{subfigure}
   \begin{subfigure}{0.495\textwidth}
       \centering
    \includegraphics[width=\columnwidth]{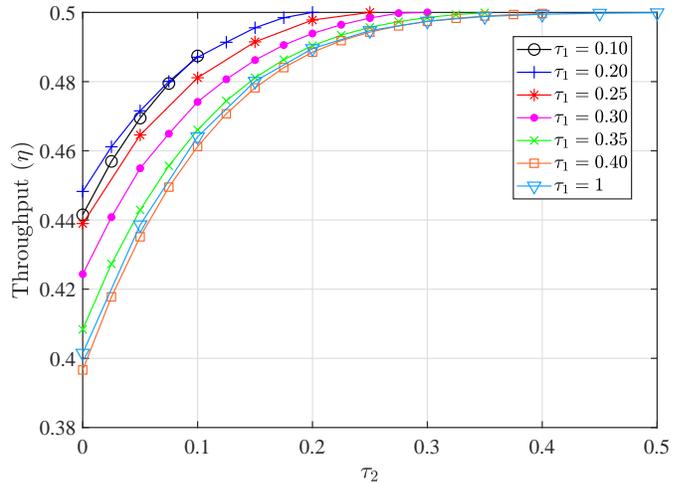}
    \caption{Throughput}
    \label{fig:PERa1a2_t}
   \end{subfigure}   
   \caption{PER and throughput performance of IR-N-HARQ in the AWGN channel when $m=3$, $k=50$, $n=100$, SNR$=-4$dB, and $\alpha_1=\alpha_2=1$.}
   \label{fig:optm2alpha}
\end{figure*}

Fig. \ref{fig:optm2alpha} shows the PER and throughput of N-HARQ for different values of time-sharing parameters when $m=3$ and $\alpha_1=\alpha_2=1$. We assume that $\tau_1\in[0, 1]$ and $\tau_2\in[0,\tau_1]$. We also assume that $\alpha_2\leq\alpha_1$ to maintain a hierarchical decoding order and indicate that the first retransmission is conducted with more power than the second retransmission.
In Fig. \ref{fig:PERa1a2_t}, we see the effect of different choices of retransmission parameters on the throughput performance of the N-HARQ. As can be seen, with $\tau=1$, the throughput gain is lower by increasing $\tau_2$. This is because when $\tau_1=1$, the retransmission slot is not sharing any packet with the new arriving packet. However, by reducing $\tau_1$ and allowing more duration to the new arriving packet,  the overall throughput is increased. It is also important to note that with lower $\tau_1$, the throughput remains higher even with a slight mismatch in the optimization parameters, while at higher values of $\tau_1$, the throughput significantly varies with a slight variation in the retransmission parameters.

\begin{figure*}[t]
\begin{subfigure}{0.495\textwidth}
    \centering
    \includegraphics[width=\columnwidth]{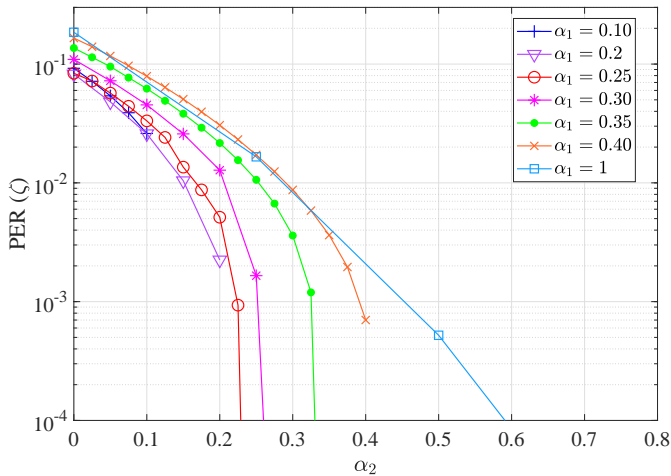}
    \caption{Packet Error Rate}
    \label{fig:PERt1t2}
\end{subfigure}
  \begin{subfigure}{0.495\textwidth}
       \centering
    \includegraphics[width=\columnwidth]{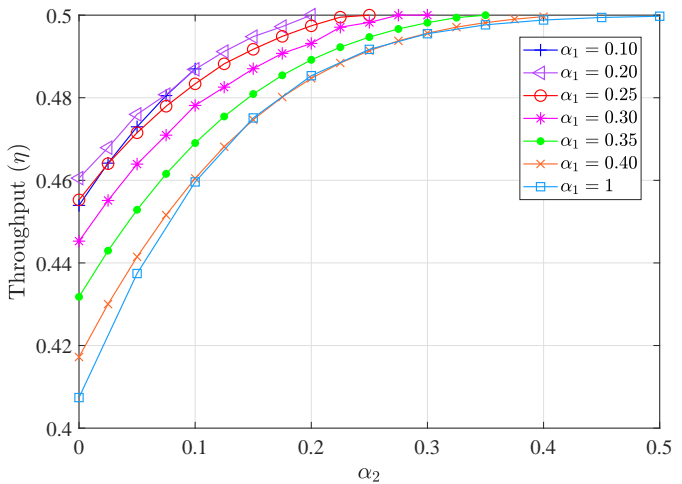}
    \caption{Throughput}
    \label{fig:PERt1t2_tp}
   \end{subfigure}   
   \caption{PER and throughput performance of IR-N-HARQ in the AWGN channel when $m=3$, $k=50$, $n=100$, SNR$=-4$dB, and $\tau_1=\tau_2=1$.}
   \label{fig:optm2tau}
   \vspace{-3ex}
\end{figure*}

The effect of power-splitting ratios $\alpha_1$ and $\alpha_2$ on the PER and throughput performance of N-HARQ is shown in Fig. \ref{fig:optm2tau}, when $m=3$ and $\tau_1=\tau_2=1$. As can be seen in Fig. \ref{fig:PERt1t2}, the PER is minimized when $\alpha_1=\alpha_2$ and they are  set to $0.20\sim0.35$. We can also see in Fig. \ref{fig:PERt1t2_tp} that the throughput is maximized when $\alpha_1$ and $\alpha_2$ are being chosen properly. For example, when $\alpha_1=0.3$, the retransmission packet is sharing the power with the new arriving packets in the entire time slot ($\tau_1=\tau_2=1$). This leads to higher throughput compared to the case when $\alpha_1=1$, where the transmission power is allocated to the retransmission packet only. 

It can be concluded that by changing time-sharing and power-splitting ratios, N-HARQ can achieve the desired level of reliability and throughput. However, one may choose to fix the time-sharing parameter to 1, i.e., $\tau_1=\tau_2=1$, and only optimize for $\alpha_1$ and $\alpha_2$. This leads to a much simpler transmitter-receiver pair, which is favorable in practice.
\subsection{Performance in Rayleigh fading }
\begin{figure*}[t]
\begin{subfigure}{0.495\textwidth}
    \centering
    \includegraphics[width=\columnwidth]{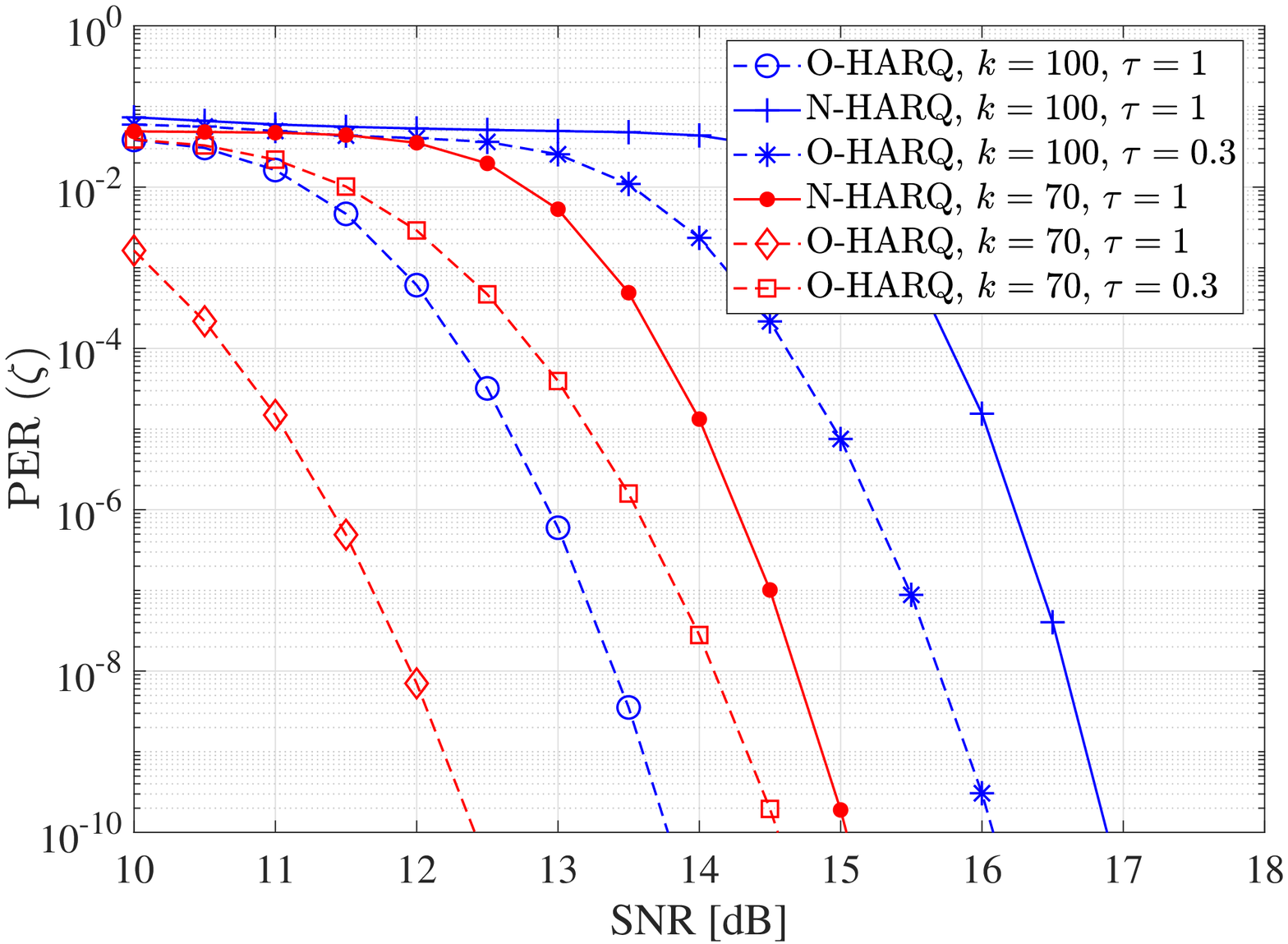}
    \caption{Packet Error Rate}
    \label{fig:PER_SNR_Ray}
\end{subfigure}
   \begin{subfigure}{0.495\textwidth}
       \centering
    \includegraphics[width=\columnwidth]{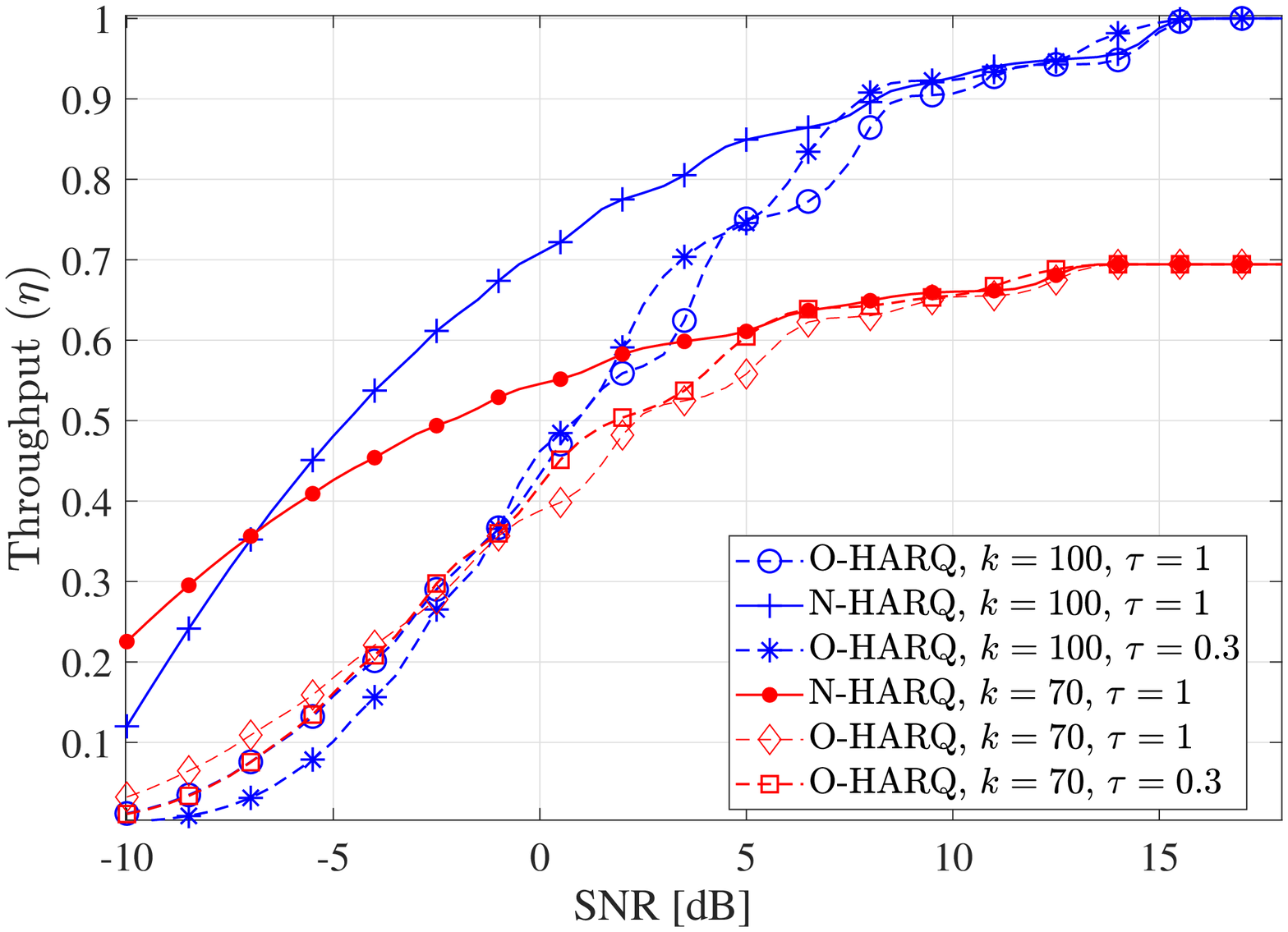}
    \caption{Throughput}
    \label{fig:th_SNR_Ray}
   \end{subfigure}   
   \caption{PER and throughput comparison between O-HARQ-IR and N-HARQ-IR, when $n=100$,  $B=100$KHz $\alpha=0.5$, and  $f_\mathrm{D}t_\mathrm{TB}=0.0338$.}
   \label{fig:Rayperth}
   \vspace{-2ex}
\end{figure*}

We assume a quasi-static channel and use FSMC to model fading by partitioning the fading envelop into $L$ fading states. We calculate fading partition so that the duration of each fading state remains fixed, where $c$ is called the channel partitioning parameter that indicates how many packets experience a fading state and therefore is a multiple of the packet length. Parameter $c$ is usually chosen to be larger than 1 and smaller than 8 \cite{sahin2019delay}.  In FSMC, the channel mobility is relative to the time block duration, i.e., a large $f_\mathrm{D}t_\mathrm{TB}$ value indicates high channel fading speed. For any fixed $c$, the number of fading states $L$ needs to be increased for decreasing $f_\mathrm{D}t_\mathrm{TB}$ value to fully capture the dynamics of the fading channel \cite{zhang1999finite}. In the rest of the paper, we assume that $c=3.0446$ and  the number of channel states $L$ is either 4 or 13. In particular, we consider two sets of channel parameters. The first scenario is when $f_\mathrm{D}=210$Hz, $t_\mathrm{TB}=0.14$ms, and $L=13$. The second scenario is when $f_\mathrm{D}=285$Hz, $t_\mathrm{TB}=0.3$ms, and $L=4$. These scenarios capture the relative velocity of $20-80$ Km/hr with carrier frequency $f_\mathrm{C}=1.9$ GHz.
\begin{figure*}[t]
    \centering
    \begin{subfigure}{.495\textwidth}
    \includegraphics[width=\columnwidth]{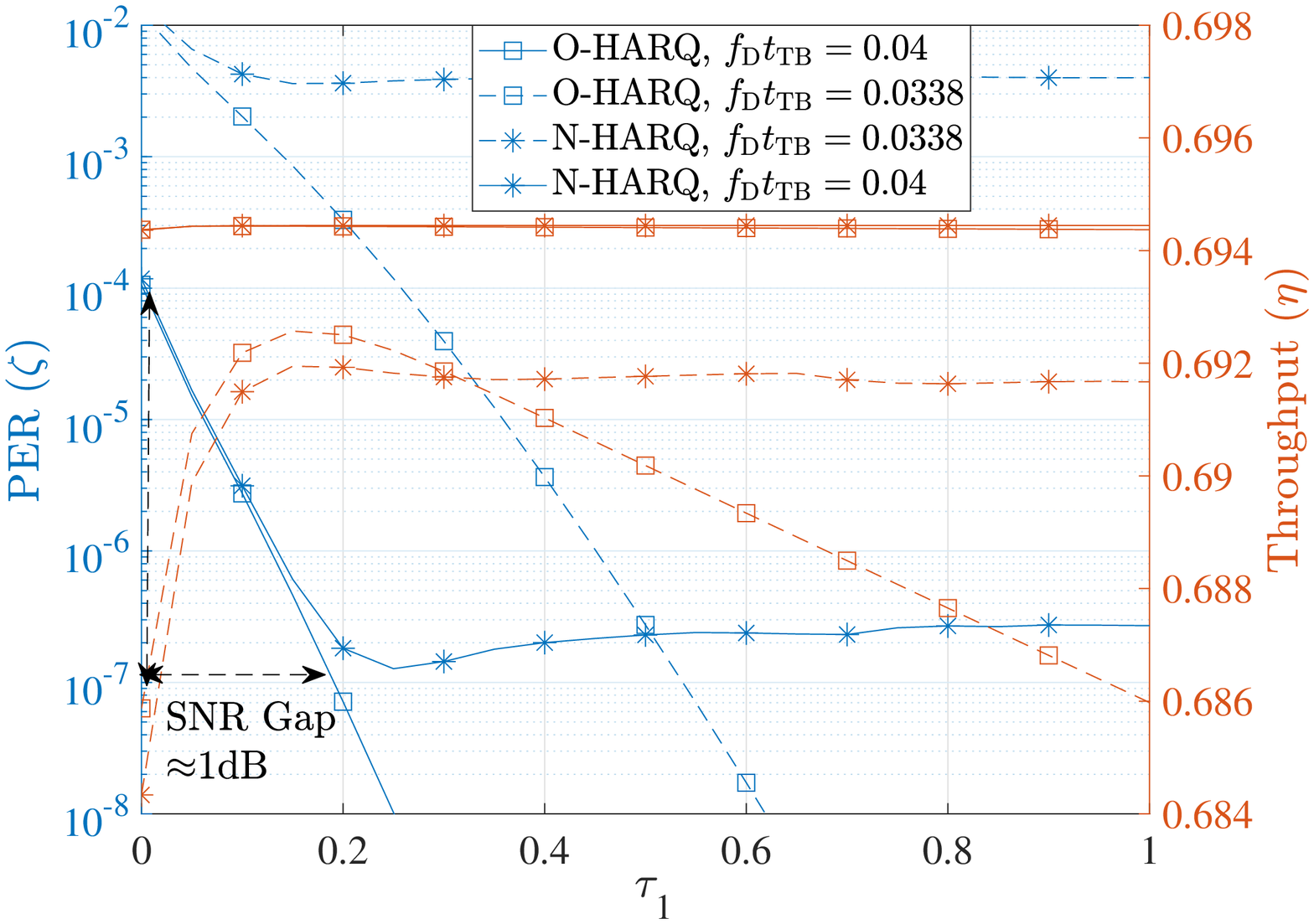}
    \caption{Reliability }
    \label{fig_delay_fd_a} 
    \end{subfigure}
    \begin{subfigure}{.495\textwidth}
    \centering
    \includegraphics[width=\columnwidth]{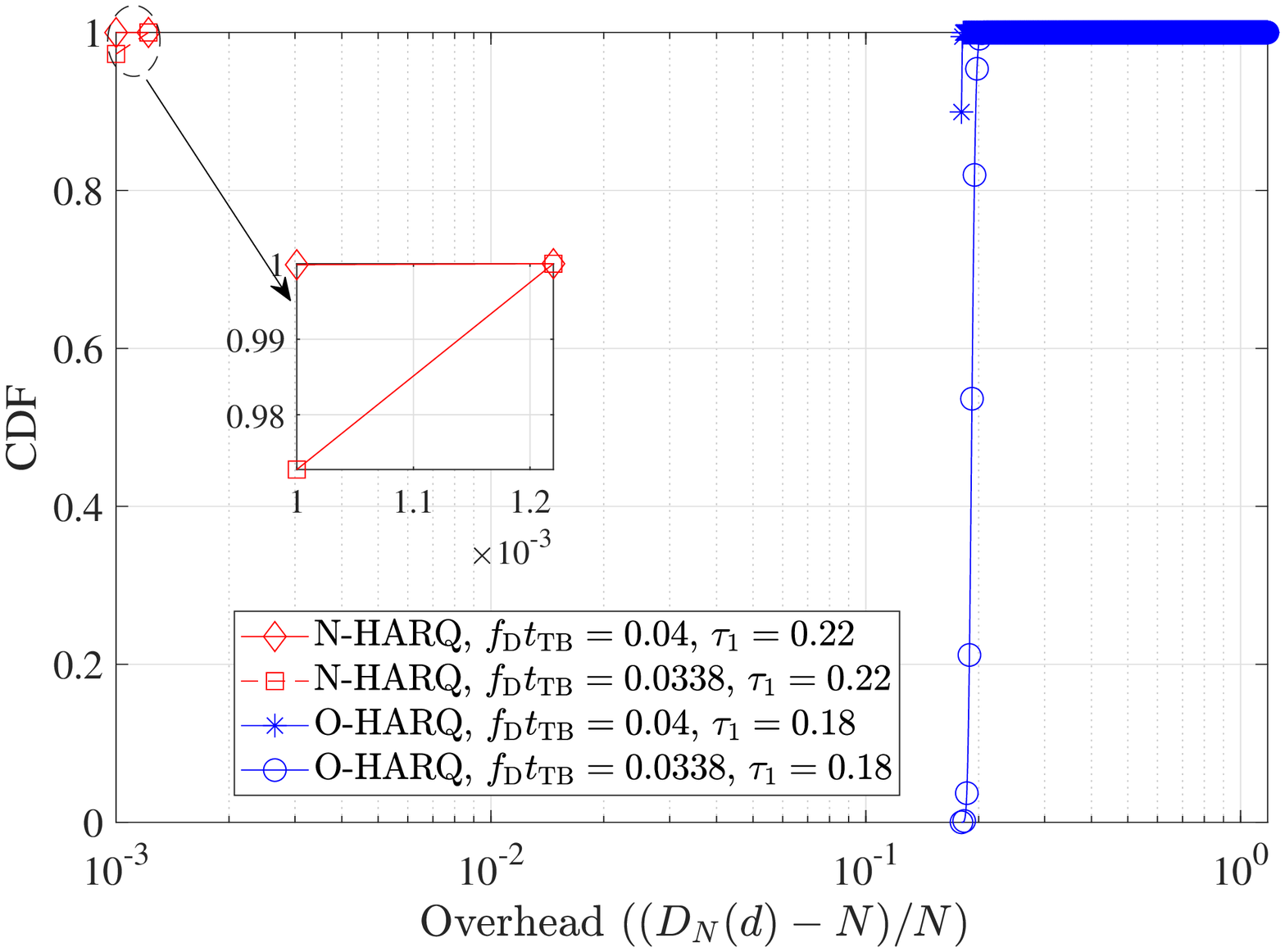}
    \caption{Latency }
    \label{fig_delay_fd_b}
    \end{subfigure}
    \caption{ Reliability and latency performance comparison between N-HARQ and O-HARQ with various relative mobility when $n=100$, $k=70$ and we choose optimal $\alpha$ for N-HARQ for each $\tau_1$. }
    \label{fig_delay_fd}
    \vspace{-3ex}
\end{figure*} 

Fig. \ref{fig:Rayperth} shows the PER and throughput of N-HARQ and O-HARQ in the fading channel for  various $\tau$ and $k$, when $m=2$. As can be seen in Fig. \ref{fig:PER_SNR_Ray}, O-HARQ outperforms N-HARQ in terms of PER across all SNRs, and there is a gap of about 3dB between N-HARQ and O-HARQ when $\tau=1$. Also, as can be seen in Fig. \ref{fig:th_SNR_Ray}, N-HARQ achieves higher throughput because it allows simultaneous transmission of the retransmission and new arriving packets. It is important to note that the fluctuations in the throughput curve are because of the finite number of channel states.

In Fig. \ref{fig_delay_fd}, the reliability and  delay performance of N-HARQ and O-HARQ in the Rayleigh fading channel are presented, when $N=1000$ packets are scheduled for transmission. Fig. \ref{fig_delay_fd_a} shows that an optimal value of $\tau_1$ can be found to minimize the PER at the target throughput in various relative mobility scenarios. While retransmission in O-HARQ achieves the desired PER and throughput performance, it has a large latency due to queuing, as shows in Fig. \ref{fig_delay_fd_b}. On the other hand, using N-HARQ, the desired PER can be achieved without causing delay overhead. For example, when $f_\mathrm{D}t_\mathrm{TB}=0.04$, the PER$\approx 10^{-7}$ can be achieved at SNR$=13$dB, by setting $\tau\approx0.18-0.2$ with O-HARQ and N-HARQ. However, the O-HARQ retransmission causes more than $10\%$ overhead in comparison to N-HARQ as shown in Fig. \ref{fig_delay_fd_b}.  An additional 1dB in the transmission power for the O-HARQ is required to compensate for the delay while achieving the PER$\approx 10^{-7}$.

\begin{figure}[t]
    \centering
    \includegraphics[width=1\columnwidth]{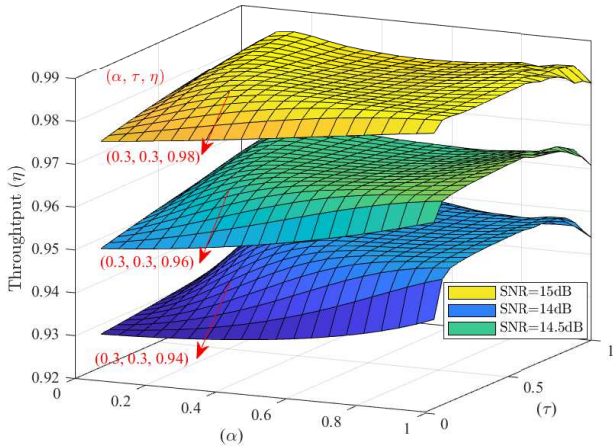}
    \caption{Throughput performance of IR-N-HARQ at various  $\alpha$ and $\tau$, when $m=1$, $n=100$, $k=100$, $t_\mathrm{TB}=0.00014$, and $f_\mathrm{D}t_\mathrm{TB}=0.0338$.  }
    \label{fig:Th_3D_Ray}
\end{figure}

Fig. \ref{fig:Th_3D_Ray} shows the effect of power-splitting ratio $\alpha$ and time-sharing parameter $\tau$, on the throughput performance of the proposed N-HARQ scheme with single retransmission in the Rayleigh fading channel. It can be seen that the throughput varies significantly with the choice of $\alpha$ and $\tau$ at a given average SNR. More specifically, when $\alpha$ is small, a larger retransmission length $\tau$ can  provide a significant reduction in the PER and accordingly, an improvement in throughput. Note that optimal points shift with the  average SNR variation. Note that points on the boundary of the surfaces are more prone to performance degradation in the event of SNR estimation error. However, there are other points that can offer good throughput with more stable performance  under SNR estimation error. This makes N-HARQ more practical and flexible from the design perspective.  

In Table \ref{TabOpt}, we list the optimal parameters that minimize PER for N-HARQ with incremental redundancy (Table \ref{Table1}) and Chase combining (Table \ref{Table2}), in the Rayleigh fading channel with different Doppler frequencies. Note that at a higher relative mobility, PER is reduced. For example, at target PER of $~10^{-5}$, there is roughly 1dB SNR gap between $f_\mathrm{D}t_\mathrm{TB}=0.0338$ and $f_\mathrm{D}t_\mathrm{TB}=0.04$. This is due to the fact that when relative mobility is high, HARQ process is more effective because of its ability to capture more time diversity of the channel. Furthermore, as it can be seen in Table \ref{Table1}, for IR-N-HARQ when  $\alpha=1$, different retransmission length $\tau$ can minimize PER at different SNRs. In general, when SNR is low, longer retransmissions are required, whereas at higher SNRs, smaller $\tau$ values minimizes the PER for both low and high mobility scenarios. As can be seen in Table \ref{Table2}, CC-N-HARQ can achieve almost the same level of PER and throughput as IR-N-HARQ when the power-splitting ratio $\alpha$ is chosen properly. Also, when SNR is in the range of low to medium  the optimization parameter $\alpha$ is small as more power is required to new arriving packets for better decoding and SIC. However, when  SNR is higher, and the receiver can decode messages with a higher probability, a larger $\alpha$ can be allocated to retransmission and small power for the new arriving packet. This may increase the percentage of packets that undergo retransmission, but due to high time-diversity, this reduces the packet error rate further.
Note that CC-N-HARQ is favorable in practice since all new and retransmission packets are sent with the same length but with different powers. This leads to a much simpler transmitter-receiver pair in comparison with IR-N-HARQ that requires variable-length packet transmission and superposition. 
\begin{table}[t]
\caption{Optimum values of $\alpha$ and $\tau$ for N-HARQ.}
\label{TabOpt}
\begin{subtable}{0.49\textwidth}
\centering
\caption{IR-HARQ}
\label{Table1}
\scriptsize
 \begin{tabular}{c|cc|cc}
\toprule
 &   \multicolumn{2}{c}{$f_\mathrm{D}t_\mathrm{TB}=0.0338$}    
 &   \multicolumn{2}{c}{$f_\mathrm{D}t_\mathrm{TB}=0.04$}    \\
  \cmidrule{2-3} \cmidrule{4-5} 
SNR & ($\hat{\alpha}$, $\hat{\tau}$) & $(\zeta, \eta)$ & ($\hat{\alpha}$, $\hat{\tau}$) & $(\zeta, \eta)$ \\
 \midrule
12   & 1, 0.40 & 0.04477, 0.94 & 1, 0.55& 0.06039, 0.93  \\ 
12.5 & 1, 0.50 & 0.04476, 0.94 & 1, 0.55  & 0.05565, 0.93 \\ 
13   & 1, 0.55 &0.03909, 0.94& 1, 0.55  & 0.03909, 0.95 \\ 
13.5 & 1, 0.55& 0.04377, 0.94& 1, 0.55  & 0.01440, 0.97\\ 
14   &1, 0.55& 0.03819, 0.95& 1, 0.15  & 0.00128, 0.99  \\ 
14.5 & 1, 0.55 &0.02304, 0.96 & 1, 0.15  &$1.7e^{-5}$ , 0.99 \\
15   & 1, 0.15&0.00617, 0.98& 1, 0.20  & $3.4e^{-5}$, 0.99\\ 
15.5 & 1, 0.15& 0.00029, 0.99& 1, 0.25  & $9.2e^{-12}$, 0.99\\ 
16   & 1, 0.20& $2.3e^{-6}$, 0.99& 1, 0.25  & $2.4e^{-16}$, 0.99   \\ 
\bottomrule
 \end{tabular}
 \end{subtable}
 \begin{subtable}{0.49\textwidth}
 \centering
\caption{CC-HARQ}
\label{Table2}
\scriptsize
 \begin{tabular}{c|cc|cc}
\toprule
 &   \multicolumn{2}{c}{$f_\mathrm{D}t_\mathrm{TB}=0.0338$}    
 &   \multicolumn{2}{c}{$f_\mathrm{D}t_\mathrm{TB}=0.04$}    \\ 
  \cmidrule{2-3} \cmidrule{4-5} 
SNR & $\hat{\alpha}$, & $(\zeta, \eta)$ & $\hat{\alpha}$ & $(\zeta, \eta)$ \\
 \midrule
12   & 0.7&   0.05476, 0.9452 &  0.40& 0.0650, 0.9349  \\ 
12.5 & 0.7 &  0.05405, 0.9459 & 0.35  & 0.061608, 0.9383\\ 
13   & 0.4 &  0.05361, 0.9463& 1  &0.053798, 0.9462 \\ 
13.5 & 0.4&  0.05012, 0.9498& 0.95  &  0.02812, 0.9718\\ 
14   & 0.35&   0.04731, 0.9526& 0.95  & 0.00611, 0.9938  \\ 
14.5 & 1 &0.03626, 0.9637& 0.95  & 0.00020,  0.9997 \\ 
15   & 1&0.01594, 0.9840& 0.95  &$ 8.7e^{-7}$,  0.9999\\ 
15.5 & 0.95& 0.00201, 0.9979& 0.95  & $5.8e^{-10}$, 0.9999\\ 
16   & 0.95& $3.4e^{-5}$, 0.9999& 0.95  & $5.0e^{-14}$, 0.9999   \\ 
\bottomrule
 \end{tabular}
 \end{subtable}
\end{table}

In summary, the proposed N-HARQ approach can be effectively used to provide per-packet delivery guarantee, which is of utmost importance for delay-sensitive applications and URLLC. N-HARQ can provide the desired level of throughput and reliability by optimizing the transmission parameters. The main advantage of N-HARQ is that it does not
 delay  new packets 
and if retransmission is needed, it can be served with the new packet simultaneously.
\subsection{Practical Consideration}
\label{Sec:Practicle_cons}
In standard O-HARQ, single-bit feedback is enough to request ACK/NACK. However, 
 in N-HARQ, the transmitter sends multiple packets in a single time slot; therefore, more bits are required to conduct proper feedback signaling. Secondly, due to non-orthogonal transmissions, the receiver complexity of N-HARQ increases, which may also increase the decoding delay because of the SIC process.  However, the decoding delay will be negligible compared with the delay associated with the retransmission, queuing, processing and feedback delay, etc. The parallel decoding techniques can significantly reduce the decoding delay related to the SIC, which can be well adapted by using  low density parity check (LDPC) codes. However, the exact delay analysis is beyond the scope of this work and would be more relevant when the actual encoder and decoder are being used\footnote{The current work can be extended to incorporate decoding delay into its analysis by adding a slight delay penalty whenever MM system transits to state $J=1$ i.e. the retransmitting state.}.

\section{Conclusion}
\label{sec:conclusion}
In this paper, we proposed a non-orthogonal HARQ (N-HARQ) strategy for delay-sensitive applications. In the proposed N-HARQ, $m$ retransmissions of the failing packet are allowed by sharing the time slot and power with new arriving packets in a non-orthogonal fashion to avoid queuing delay. Using the Marov model, the performance of N-HARQ in the FBL regime in terms of reliability, throughput, and delay was evaluated in the AWGN and Rayleigh fading channels. We also proposed an optimization framework to find the optimal time-sharing and power-splitting parameters in order to minimize the packet error rate at the desired level of throughput. Simulation results demonstrated that the proposed N-HARQ can be finely tuned to meet the desired level of reliability and throughput to guarantee the packet delivery delay. We also demonstrated that N-HARQ is superior to the conventional HARQ scheme in achieving the target reliability with guaranteed delivery delay per packet.

\appendices
% \appendix
\section{Proof of Lemma \ref{lem:Pim3awgn}}
\label{app:prooflemmam3}
\subsection{When the system transits form states $i\in\{0,1\}$}
When the system transits from state $J=i$, for $i\in\{0,1\}$, to state $J=0$ and $J=1$, the packet is successfully delivered without or with single retransmission, respectively. The state transition probability can be easily obtained similar to the proof of Lemma \ref{lem:AWGN_pij_m2}. The state transit from $J=i$, for $i\in\{0,1\}$, to state $J=2$ when decoding is successful. This happens with probability $\epsilon_{\mathrm{ir}}([\gamma_i,\gamma_0,\gamma_{I_1}], [\tau_1,{1-\tau_1},{\tau_1}])-\epsilon_{\mathrm{ir}}([\gamma_i,\gamma_0,\gamma_{I_1},\gamma_{I_2}], [\tau_1,{1-\tau_1},{\tau_1}, \tau_2])$, otherwise it transitions to $J=e$ with probability $ \epsilon_{\mathrm{ir}}([\gamma_i,\gamma_0,\gamma_{I_1},\gamma_{I_2}], [\tau_1,{1-\tau_1},{\tau_1}, \tau_2])$ following similar arguments as provided in the proof of Lemma \ref{lem:AWGN_pij_m2}.

\subsection{When the system transits form states $i\in\{2,e\}$}
\begin{figure}[t]
    \centering
    \includegraphics[width=0.8\columnwidth]{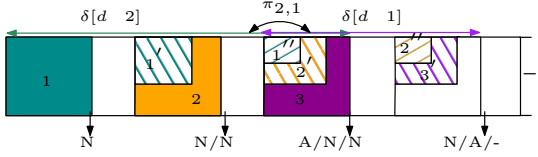}
    \caption{Example of the event $\mathcal{B}_{1}(k)$ when state transition form ${i}=2$ to $j=1$ given $k=e$. }
    \label{fig:app_example}
    \vspace{-1em}
\end{figure}

When the system state  transit from state $J=i$ for $i\in\{2,e\}$  to $J=j$, an additional packet  in state  $J=k$ can be presented in between the state transition which can influence the state transition probabilities.  This is due to the fact that in the Markov model, states are defined in terms of their delay overhead. Therefore due to non-orthogonal scheduling, when the state transit from $J=i$, state transition boundary between two packets can occurs during last retransmission of the first  packet and the first transmission of the third packet. As shown in Fig. \ref{fig:app_example}, packet number 1 is in state $i=2$ and  its boundary is overlapping with the first transmission of  packet  number 3, so packet number 2 is an additional packet and this event is denoted as $\mathcal{B}_{r}(k)$, where $r\in\{0,1,2\}$ represents the retransmission number.
Let $p_k\mathrm{Prob}\{\mathcal{B}_0(k)\}$, $p_k\mathrm{Prob}\{\mathcal{B}_1(k) \}$ and $p_k\mathrm{Prob}\{\mathcal{B}_2(k)\}$ represents error probabilities of a packet with no retransmission, with single retransmission, and with 2 retransmissions, respectively, where $p_k$ is the marginal probability of the packet in state $J=k$ and $\sum_kp_k=1$. The system transits from state $J=i$ to state $J=0$ with an additional packet in state $J=k$ with probability   $\sum_k p_k(1-\mathrm{Prob}\{\mathcal{B}_0(k)\})$, where the total state transition probability is  accumulated for all $J=k$. The proof follows similar as proof of Lemma \ref{lem:AWGN_pij_m2}. 
Similarly,  the system transits from state $J=i$ to states $J\in\{1,2\}$, when  the packet is successful with single or two retransmissions with probability $\sum_kp_k(\mathrm{Prob}\{\mathcal{B}_0(k)\}- \mathrm{Prob}\{\mathcal{B}_1(k)\} )$ and  $\sum_kp_k(\mathrm{Prob}\{\mathcal{B}_1(k)\}- \mathrm{Prob}\{\mathcal{B}_2(k)\} )$; otherwise, it transits to states $J=e$  with probability $\sum_kp_k( \mathrm{Prob}\{\mathcal{B}_2(k)\} )$,  following similar approach as proof of Lemma \ref{lem:AWGN_pij_m2}. 

When $i=2$, the packet interference can be removed and the SINR can be increased, where as with $i=e$ it cannot be decoded. Similarly, each $k$-th state of the intermediate packet can influence the SINR resulting in change of state transition probabilities. The probabilities of events  $\mathcal{B}_r(k)$ ($r\in\{0,1,2\}$) for each $r$ and $k$ is given as follows: 
\subsubsection{When $i=2$} Firstly, we have:
\begin{align}
\nonumber\mathrm{Prob}\{\mathcal{B}_0(0)\}&=\epsilon_\mathrm{ir}([\gamma_{\bar{I}_2}, \gamma_0],[\tau_2, \bar{\tau}_2]),\\ \nonumber\mathrm{Prob}\{\mathcal{B}_1(0)\}&=\epsilon_\mathrm{ir}([\gamma_{\bar{I}_2}, \gamma_0,\gamma_{I_1}],[\tau_2, \bar{\tau}_2, \tau_1]),\\ \nonumber\mathrm{Prob}\{\mathcal{B}_2(0)\}&=\epsilon_\mathrm{ir}([\gamma_{\bar{I}_2}, \gamma_0,\gamma_{I_1},\gamma_{I_2}],[\tau_2, \bar{\tau}_2, \tau_1,\tau_2]),
\end{align}
where $\gamma_{\bar{I}_2}=(1-\alpha_2)P$, $\gamma_0=P$ $\gamma_{I_1}=\frac{\alpha_1 P}{1+(1-\alpha_1)P }$, $\gamma_{I_2}=\frac{P\alpha_2}{1+(1-\alpha_2)P}$, and $\bar{\tau}_2 = 1- \tau_2$. 
Secondly, when $k=1$, we have:
\begin{align}
\nonumber\mathrm{Prob}\{\mathcal{B}_0(1)\}&=\epsilon_\mathrm{ir}([\gamma_{\bar{I}_1}, \gamma_0],[\tau_1, \bar{\tau}_1]),\\ \nonumber\mathrm{Prob}\{\mathcal{B}_1(1)\}&=\epsilon_\mathrm{ir}([\gamma_{\bar{I}_1}, \gamma_0,\gamma_{I_1}],[\tau_1, \bar{\tau}_1, \tau_1]),\\
\nonumber\mathrm{Prob}\{\mathcal{B}_2(1)\}&=\epsilon_\mathrm{ir}([\gamma_{\bar{I}_1}, \gamma_0,\gamma_{I_1},\gamma_{I_2}],[\tau_1, \bar{\tau}_1, \tau_1, \tau_2]),
\end{align}
where $\gamma_{\bar{I}_1}  = (1-\alpha_1)P$ and $\bar{\tau}_1=1-\tau_1$.
Thirdly,
when $k=2$, we have: \begin{align}
\nonumber\mathrm{Prob}\{\mathcal{B}_0(2)\}&=\epsilon_\mathrm{ir}([\gamma_{\bar{I}_1}, \gamma_0],[\tau_1, \bar{\tau}_1]),\\
\nonumber\mathrm{Prob}\{\mathcal{B}_1(2)\}&=\epsilon_\mathrm{ir}([\gamma_{\bar{I}_1}, \gamma_0, \gamma_{\bar{I}_{1a}}, \gamma_{\bar{I}_1}],[\tau_1, \bar{\tau}_1, \tau_2, \tau_{12}]),\\
\nonumber\mathrm{Prob}\{\mathcal{B}_2(2)\}&=\epsilon_\mathrm{ir}([\gamma_{\bar{I}_1}, \gamma_0, \gamma_{{I}_{1a}}, \gamma_{{I}_1}, \gamma_{{I}_2}],[\tau_1, \bar{\tau}_1, \tau_2, \tau_{12}, \tau_2]),
\end{align}
where $\tau_{12}=\tau_{1}-\tau_2$ and $\gamma_{\bar{I}_{1a}}= \frac{(\alpha_1-\alpha_2)P}{1+(1-\alpha_1)P}$.
Finally,
when $k=e$, we have: \begin{align}
\nonumber\mathrm{Prob}\{\mathcal{B}_0(e)\}&=\epsilon_\mathrm{ir}([\gamma_{\bar{E}_{1a}},  \gamma_{\bar{E}_{1}}, \gamma_0],[\tau_2, \tau_{12}, \bar{\tau}_1]),\\ \nonumber\mathrm{Prob}\{\mathcal{B}_1(e)\}&=\epsilon_\mathrm{ir}([\gamma_{\bar{E}_{1a}},  \gamma_{\bar{E}_{1}}, \gamma_0, \gamma_{E_1}, \gamma_{{I}_1}],[\tau_2, \tau_{12}, \bar{\tau}_1, \tau_2, \tau_{12}]),\\
\nonumber\mathrm{Prob}\{\mathcal{B}_2(e)\}&=\\
\nonumber\epsilon_\mathrm{ir}([&\gamma_{\bar{E}_{1a}},\gamma_{\bar{E}_{1}} , \gamma_0, \gamma_{{E}_1}, \gamma_{{I}_1}, \gamma_{{I}_2}],[\tau_2, \tau_{12}, \bar{\tau}_1, \tau_2, \tau_{12}, \tau_2]),
\end{align}
where 
$\gamma_{\bar{E}_{1a}}= \frac{(1-\alpha_1)P}{1+(\alpha_1-\alpha_2)P}$, $\gamma_{\bar{E}_1}  = \frac{(1-\alpha_1)P}{1+\alpha_1P}$ and $\gamma_{E_1}= \frac{(\alpha_1-\alpha_2)P}{1+(1-(\alpha_1-\alpha_2))P}$.
\subsubsection{When $i=e$}
Firstly, 
when $k=0$, we have:
\begin{align}
\nonumber\mathrm{Prob}\{\mathcal{B}_0(0)\}&=\epsilon_\mathrm{ir}([\gamma_{\bar{E}_2}, \gamma_0],[\tau_2, \bar{\tau}_2]),\\  \nonumber\mathrm{Prob}\{\mathcal{B}_1(0)\}&=\epsilon_\mathrm{ir}([\gamma_{\bar{E}_2}, \gamma_0, \gamma_{{I}_1}],[\tau_2, \bar{\tau}_2, \tau_1]),\\ \nonumber\mathrm{Prob}\{\mathcal{B}_2(0)\}&=\epsilon_\mathrm{ir}([\gamma_{\bar{E}_2}, \gamma_0, \gamma_{{I}_1}, \gamma_{{I}_2}],[\tau_2, \bar{\tau}_2, \tau_1, \tau_2]),
\end{align}
where
$\gamma_{\bar{E}_2}=\frac{(1-\alpha_2)P}{1+\alpha_2 P}$.
Secondly, 
when $k=1$, we have:
\begin{align}
\nonumber\mathrm{Prob}\{\mathcal{B}_0(1)\}&=\epsilon_\mathrm{ir}([\gamma_{\bar{E}_{2a}}, \gamma_{\bar{I}_1}, \gamma_0],[\tau_2, \tau_{12}, \bar{\tau}_1]),\\
\nonumber\mathrm{Prob}\{\mathcal{B}_1(1)\}&=\epsilon_\mathrm{ir}([\gamma_{\bar{E}_{2a}}, \gamma_{\bar{I}_1}, \gamma_0, \gamma_{{I}_1}],[\tau_2, \tau_{12}, \bar{\tau}_1, \tau_1]),\\ \nonumber \mathrm{Prob}\{\mathcal{B}_2(1)\}&=\epsilon_\mathrm{ir}([\gamma_{\bar{E}_{2a}}, \gamma_{\bar{I}_1}, \gamma_0, \gamma_{{I}_1}, \gamma_{{I}_2}],[\tau_2, \tau_{12}, \bar{\tau}_1, \tau_1, \tau_2]),
\end{align}
where
$\gamma_{\bar{E}_{2a}}=\frac{(1-\alpha_1)P}{1+\alpha_2P}$. Thirdly,
when $k=2$, we have:
\begin{align}
\nonumber\mathrm{Prob}\{\mathcal{B}_0(2)\}&=\epsilon_\mathrm{ir}([\gamma_{\bar{E}_{2a}}, \gamma_{\bar{I}_1}, \gamma_0],[\tau_2, \tau_{12}, \bar{\tau}_1]),\\
\nonumber\mathrm{Prob}\{\mathcal{B}_1(2)\}&=\epsilon_\mathrm{ir}([\gamma_{\bar{E}_{2a}}, \gamma_{\bar{I}_1}, \gamma_0,\gamma_{\bar{I}_{1a}}, \gamma_{{I}_{1}} ],[\tau_2, \tau_{12}, \bar{\tau}_1, \tau_2, \tau_{12}]),\\
\nonumber\mathrm{Prob}\{\mathcal{B}_2(2)\}&=\\
\nonumber\epsilon_\mathrm{ir}([&\gamma_{\bar{E}_{2a}}, \gamma_{\bar{I}_1}, \gamma_0,\gamma_{\bar{I}_{1a}}, \gamma_{{I}_{1}}, \gamma_{{I}_{2}} ],[\tau_2, \tau_{12}, \bar{\tau}_1, \tau_2, \tau_{12}, \tau_2]).
\end{align}
Finally, when $k=e$, we have: \begin{align}
\nonumber\mathrm{Prob}\{\mathcal{B}_0(e)\}&=\epsilon_\mathrm{ir}([\gamma_{\bar{E}_{1}},  \gamma_0],[\tau_1, \bar{\tau}_1]),\\ \nonumber\mathrm{Prob}\{\mathcal{B}_1(e)\}&=\epsilon_\mathrm{ir}([\gamma_{\bar{E}_{1}},  \gamma_0, \gamma_{{E}_{1}},\gamma_{{I}_{1}} ],[\tau_1, \bar{\tau}_1, \tau_2, \tau_{12}]),\\ \nonumber\mathrm{Prob}\{\mathcal{B}_2(e)\}&=\epsilon_\mathrm{ir}([\gamma_{\bar{E}_{1}},  \gamma_0, \gamma_{{E}_{1}},\gamma_{{I}_{1}} , \gamma_{{I}_{2}}],[\tau_1, \bar{\tau}_1, \tau_2, \tau_{12}, \tau_2]).
\end{align}

\footnotesize
\bibliographystyle{IEEEtran}
\bibliography{main.bib}
\vspace{-1cm}
\begin{IEEEbiography}[{\includegraphics[width=1in,height=1.25in,clip,keepaspectratio]{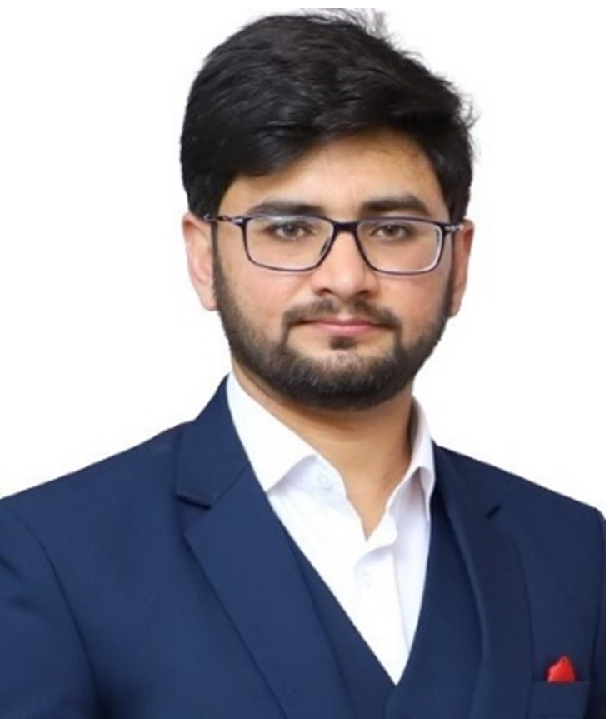}}]%
{Faisal Nadeem}
(Student Member IEEE) received the M.Phil degree in Communication and Signal Processing from the Department of Electronics, Quaid-i-Azam University, Islamabad, Pakistan in 2014.  He is currently pursuing the Ph.D. degree with the School of Electrical and Information Engineering, The University of Sydney, Australia. His research interests lies in ultrareliable low-latency communications and massive machine type communication for industrial Internet of Things (IIoT). 
\end{IEEEbiography}
\vspace{-3cm}
\begin{IEEEbiography}
[{\includegraphics[width=1in,height=1.25in,clip,keepaspectratio]{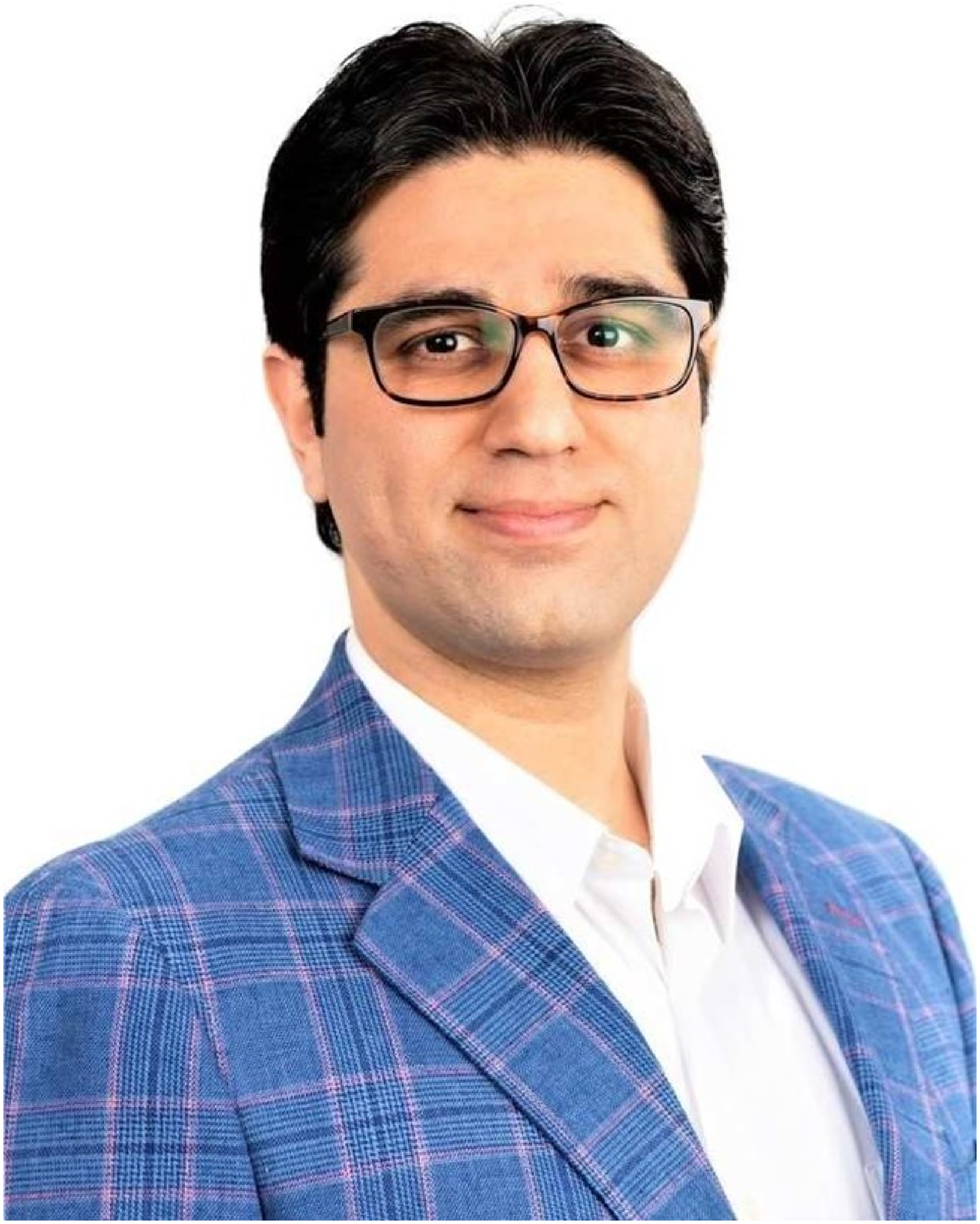}}]
{Mahyar Shirvanimoghaddam}
 (Senior Member, IEEE) received the B.Sc. degree (Hons.) from the University of Tehran, Iran, in 2008, the M.Sc. degree (Hons.) from the Sharif University of Technology, Iran, in 2010, and the Ph.D. degree from The University of Sydney, Australia, in 2015, all in electrical engineering. He held a post-doctoral research position at the School of Electrical Engineering and Computer Science, University of Newcastle, Australia. Since 2016, he has been with the School of Electrical and Information Engineering, The University of Sydney, as an Academic Fellow of telecommunications. In 2018, he was named as one of the top 50 Young Scientists under the age of 40 by the World Economic Forum for his contributions to the Internet of Things technologies and Industry 4.0. His general research interests include channel coding techniques, massive multiple access, and communication strategies for the internet of things. He is the Exemplary Reviewer of the IEEE Transactions on Communications from 2016 to 2019 and IEEE Communications Letters in 2016.
\end{IEEEbiography}
\vspace{-3cm}
\begin{IEEEbiography}[{\includegraphics[width=1in,height=1.25in,clip,keepaspectratio]{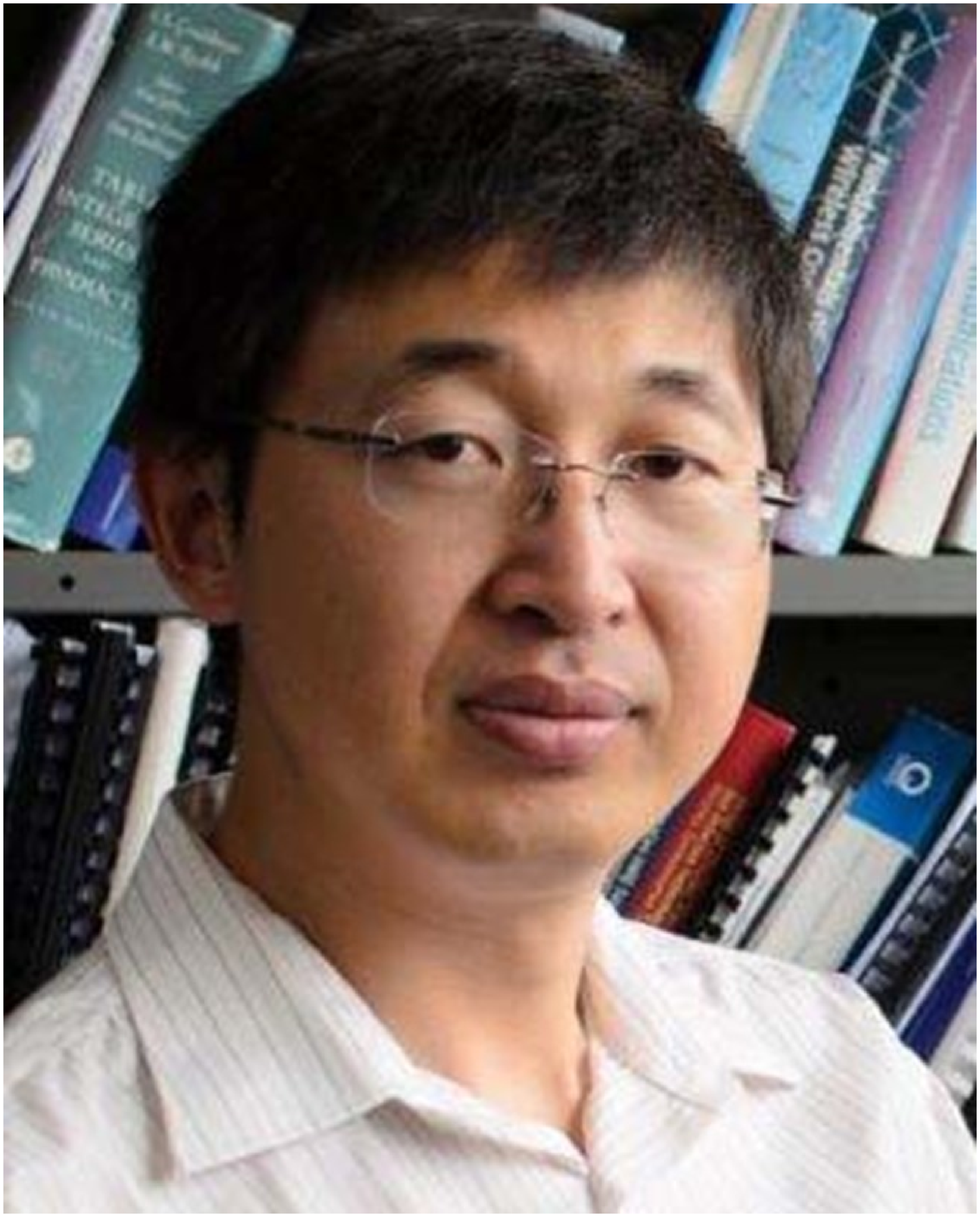}}]%
{Yonghui Li }
(Fellow, IEEE) received the Ph.D. degree from the Beijing University of Aeronautics and Astronautics, in November 2002. From 1999 to 2003, he was affiliated with Linkair Communication Inc., where he held a position of project manager with responsibility for the design of physical layer solutions for the LAS-CDMA systems. Since 2003, he has been with the Centre of Excellence in Telecommunications, The University of Sydney, Australia. He is currently a Professor with the School of Electrical and Information Engineering, The University of Sydney. His current research interests are in the area of wireless communications, with a particular focus on MIMO, millimeter wave communications, machine to machine communications, coding techniques, and cooperative communications. He holds a number of patents granted and pending in these fields. He was a recipient of the Australian Queen Elizabeth II Fellowship in 2008 and the Australian Future Fellowship in 2012. He is also an Editor for the IEEE Transactions on Communications and the IEEE Transactions on Vehicular Technology. He also served as a Guest Editor for several special issues of IEEE journals, such as the IEEE JSAC special issue on Millimeter Wave Communications. He received the best paper awards from the IEEE International Conference on Communications (ICC) 2014, the IEEE PIMRC 2017, and the IEEE Wireless Days Conferences (WD) 2014.
\end{IEEEbiography}
\vspace{-3cm}
\begin{IEEEbiography}[{\includegraphics[width=1in,height=1.25in,clip,keepaspectratio]{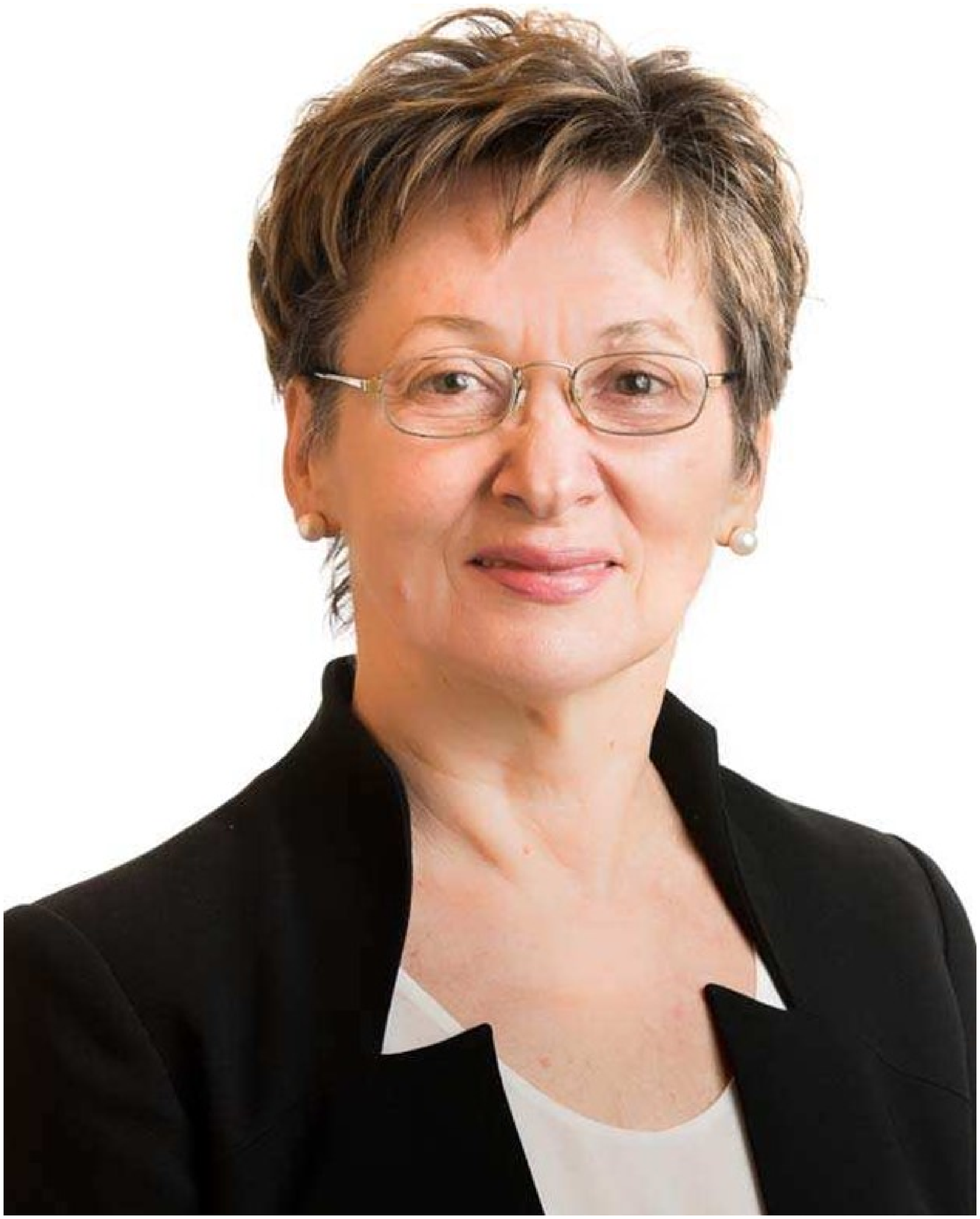}}]%
{Branka Vucetic }
 (Life Fellow, IEEE) received the B.S.E.E., M.S.E.E., and Ph.D. degrees in electrical engineering from The University of Belgrade, Belgrade, in 1972, 1978, and 1982, respectively. She is an ARC Laureate Fellow and the Director of the Centre of Excellence for IoT and Telecommunications, The University of Sydney. Her current work is in the areas of wireless networks and the Internet of Things. In the area of wireless networks, she explores ultrareliable, low-latency techniques, and transmission in millimetre wave frequency bands. In the area of the Internet of things, Vucetic works on providing wireless connectivity for mission critical applications. She is a fellow of the Australian Academy of Science and the Australian Academy of Technological Sciences and Engineering.
\end{IEEEbiography}

\end{document}